\documentclass[12pt]{article}
\usepackage[utf8]{inputenc}
\usepackage{amsmath,amssymb}
\usepackage{graphicx}
\usepackage{array}
\usepackage{booktabs}
\usepackage{hyperref}

\def \lbdd {\overline{\Lambda}^{\ddagger}}

\newcommand{\vf}{\varphi}

\newcommand{\be}{\begin{equation}}
	\newcommand{\ee}{\end{equation}}
\def\bea{\begin{eqnarray}}
	\def\eea{\end{eqnarray}}
\newcommand{\nn}{\nonumber}
\newcommand{\no}{\noindent}
\newcommand{\p}{\partial}

\def \olamb {\overline \Lambda }

\begin{document}
	
		\title{\textbf{Minimal massive completions of partially broken higher-spin models}}
	
	\author{  E. L. Mendonça\footnote{elias.leite@unesp.br} , M. I. C. Maués \footnote{isabel.maues@unesp.br} . \\
		\textit{Departamento de Física, FEG-UNESP}\\
		\textit{Av. Dr. Ariberto Pereira da Cunha, 333, Guaratinguetá, SP, Brazil.}}
	\date{\today}
	\maketitle

	\begin{abstract}
		The partially broken Fronsdal (PBF) model describes massless particles of arbitrary integer spin $s$ with a gauge symmetry smaller than Fronsdal's, and differs from it only by the coefficient $c$ of the term $(\p\cdot h')^2$. We ask whether it admits a genuinely new massive completion. Adding a mass term and the most general local auxiliary sector with at most two derivatives, and demanding that all lower-spin sectors be non-dynamical, we show that for $s=3$ a single auxiliary scalar is never enough, while two scalars give a solution unique up to field redefinitions: one scalar is algebraic and its elimination exactly undoes the deformation, while the other is the auxiliary scalar of the Aragone--Deser--Yang (ADY) model. The same mechanism, with an algebraic vector, appears for $s=4$ and extends to arbitrary spin, and we conjecture that the resulting auxiliary content is minimal. In this minimal completion $c$ does not survive the mass: the correlation functions of $h$ coincide with those of the ADY model. For the Skvortsov--Vasiliev (SV) model, which belongs to the same family, we construct a massive version in which the Weyl-like symmetry survives the mass, and show that its auxiliary sector is fixed by the absence of lower-spin modes, within a general ansatz for $s=3$ and within a restricted one for $s=4$. For the partially broken SV model we obtain instead a massive completion which requires no auxiliary field beyond those of the massive SV model, and in which $c$ survives the mass through the same contact term as in the massless theory.
	\end{abstract}
	
	\newpage
		\section{Introduction}
	
	The description of free massive particles of arbitrary integer spin by local Lagrangians is an old problem. For a massive spin-$s$ particle described by a rank-$s$ tensor, the Fierz--Pauli conditions \cite{fp} cannot be derived from a Lagrangian for $s\geq3$ unless auxiliary fields are introduced. Singh and Hagen (SH) \cite{sh,sh2} solved the problem with a tower of auxiliary fields of lower rank, and in the massless limit the rank-$s$ and rank-$(s-2)$ fields combine into the double traceless field of the Fronsdal model \cite{fronsdal}. Aragone, Deser and Yang (ADY) \cite{ady} showed that this structure follows from the Kaluza--Klein reduction of the massless Fronsdal theory in $D+1$ dimensions, which also gives a Stückelberg formulation of the massive theory \cite{zinoviev}. For reviews on higher spin fields see \cite{vasiliev_r,st_r,rt,snow,book,vasilievr2}.
	
	In all these constructions the gauge symmetry of the Fronsdal Lagrangian plays an essential role: its Noether identity turns the divergence of the equations of motion into a constraint as soon as a mass term is added. It is therefore natural to ask what happens when the kinetic term is replaced by one describing the same massless particle with a smaller gauge symmetry. A concrete example is the partially broken Fronsdal (PBF) model \cite{pbf}, which differs from Fronsdal's only by the coefficient $c$ of the term $(\p\cdot h')^2$ and whose gauge parameter is restricted by a vanishing double divergence. It propagates only massless spin-$s$ modes \cite{pbf}, and its two-point amplitudes coincide with the Fronsdal ones up to a contact term proportional to $1/(c-1)$ \cite{dmmb}. The same family of Lagrangians contains a second line of models: the Skvortsov--Vasiliev (SV) model \cite{sv}, invariant under transverse diffeomorphisms and Weyl-like transformations, and its partially broken version (PBSV) \cite{dmmb}. See also \cite{tdiffspin3} for related models.
	
	In this work we ask whether these partially broken models admit their own massive completions, and which auxiliary fields such completions require. Following the ADY philosophy, we start from the massless Lagrangian, add a mass term and the most general local auxiliary sector with at most two derivatives, and fix all the coefficients by requiring, sector by sector in spin, that only a massive spin-$s$ particle propagates. The auxiliary content is not assumed in advance, and it is enlarged only when smaller sets are proven to fail.
	
	For the PBF model we find that, for $s=3$, a single auxiliary scalar is never sufficient when $c\neq1$, while two scalars lead to a solution which is unique up to field redefinitions: one scalar is algebraic and its elimination exactly cancels the deformation, while the other one is the ADY scalar. For $s=4$ the ADY vector and scalar are supplemented by one algebraic vector, and for every $s$ and $D$ the PBF model admits a massive completion with the ADY content plus one algebraic traceless field of rank $s-3$. We conjecture that this auxiliary content is minimal for every $s$; we prove it for $s=3$ and show it for $s=4$ within the branches we follow. In this minimal completion the parameter $c$ disappears from the correlation functions of $h$ and of the ADY auxiliary fields, which coincide with those of the ADY model for every $m\neq0$.
	
	For the SV line the outcome is different. The SV Lagrangian is the Fronsdal Lagrangian of the traceless part of $h$, so the trace, which in the ADY construction plays the role of the rank-$(s-2)$ auxiliary field of SH, is pure gauge. Reinstating it as an independent field, we obtain a massive SV model, equivalent to ADY, in which the Weyl-like symmetry survives the mass; its auxiliary sector is fixed by the absence of lower-spin modes, within a general ansatz for $s=3$ and within a restricted one for $s=4$. A massive PBSV model is then obtained with no auxiliary field beyond those of the massive SV model, since the deformation acts on a decoupled sector, and in it $c$ survives the mass through the same contact term as in the massless theory; adding the algebraic field of the PBF construction restores the full Weyl symmetry and removes $c$.
	
	The paper is organized as follows. Section~\ref{sec:pbf} presents the family $\mathcal{L}^{(s)}(a,b,c)$ and the PBF and PBSV models, and Section~\ref{sec:shdy} reviews the SH and ADY models and the spin projection criterion. Section~\ref{sec:spin3} contains the massive PBF models for $s=3$ and $s=4$, and Section~\ref{sec:spins} their extension to arbitrary spin. Sections~\ref{sec:sv} and \ref{sec:pbsv} are devoted to the massive SV and PBSV models, and Section~\ref{sec:conclusion} to the correlation functions and open questions. Our notation\footnote{We work in flat Minkowski spacetime with $\eta=\mathrm{diag}(-,+,\dots,+)$ and $D\ge 3$ dimensions. We adopt the compact notation of \cite{fs1}, summarized in footnote~\ref{fn:notation}. The label $T$ indicates a transverse object and $TT$ a transverse and traceless one.} is as in \cite{dmmb}.
\section{Massless models and their partially broken partners}\label{sec:pbf}

We consider the three-parameter family of second-order Lagrangians of \cite{pbf} for a double traceless ($h''=0$) totally symmetric rank-$s$ field,\footnote{Indices are not written explicitly. A dot denotes the contraction of one pair of indices and a prime a trace; for instance, $(\p\cdot h)_{\mu_2\dots\mu_s}=\p^{\mu}h_{\mu\mu_2\dots\mu_s}$, $(\p\cdot\p\cdot h)_{\mu_3\dots\mu_s}=\p^{\mu}\p^{\nu}h_{\mu\nu\mu_3\dots\mu_s}$ and $h'_{\mu_3\dots\mu_s}=\eta^{\mu\nu}h_{\mu\nu\mu_3\dots\mu_s}$. A gradient $\p X$ or a product $\eta X$ denotes the symmetrized expression with the minimal number of terms and without weights; for instance, $\p\olamb$ contains $s$ terms, $\p_{\mu_1}\olamb_{\mu_2\dots\mu_s}+\dots$, and $\eta\,h'$ contains $s(s-1)/2$ terms. Quadratic expressions in two tensors of the same rank are full contractions with unit weight, $X\cdot Y=X_{\mu_1\dots\mu_r}Y^{\mu_1\dots\mu_r}$. In (\ref{labc}), $h\,\Box\,h=h\cdot\Box h$, $(\p\cdot h)^2=(\p\cdot h)\cdot(\p\cdot h)$, $\p\cdot\p\cdot h\;h'=(\p\cdot\p\cdot h)\cdot h'$ and $(\p\cdot h')^2=(\p\cdot h')\cdot(\p\cdot h')$, while $(\p h')^2=\p_\rho h'_{\mu_3\dots\mu_s}\,\p^\rho h'^{\mu_3\dots\mu_s}$ involves no symmetrization.\label{fn:notation}}
\be
\mathcal{L}^{(s)}(a,b,c) = \frac{1}{2}\,h\,\Box\, h + \frac{s}{2}\,(\p\cdot h)^2 + \frac{a\, s(s-1)}{2}\, \p\cdot\p\cdot h\; h' + \frac{b\, s(s-1)}{4}\,(\p h')^2 + \frac{3\, c}{4}\binom{s}{3}(\p\cdot h')^2 \, , \label{labc}
\ee
which, for all $(a,b,c)\in\mathbb{R}$, is invariant up to total derivatives under traceless and transverse higher-spin diffeomorphisms, $\delta h=\p\,\olamb^T$, which remove the spin-$(s-1)$ modes \cite{pbf,dmmb}. For $a\neq a^*_s\equiv2/[D+2(s-2)]$, the local and invertible field redefinition ($r$-shift)
\be
h\to h+r\,\eta\, h'\, ,\qquad r=\frac{1-a}{a[D+2(s-2)]-2}\, , \label{rshift}
\ee
brings (\ref{labc}) into the form \cite{dmmb}
\be
\mathcal{L}^{(s)}(a,b,c)=\mathcal{L}^{(s)}_F(h)+\frac{s(s-1)(D+2s-6)}{4\,[a(D+2s-4)-2]^2}\Big[f^{(s)}_b\,h'\,\Box\,h'+\frac{s-2}{2}\,f^{(s)}_c\,(\p\cdot h')^2\Big]\, , \label{lsplit}
\ee
where $\mathcal{L}^{(s)}_F\equiv\mathcal{L}^{(s)}(1,1,1)$ is the Fronsdal Lagrangian \cite{fronsdal} and
\be
f^{(s)}_b=(D+2s-6)(a^2-b)+(a-1)^2\, ,\qquad f^{(s)}_c=(D+2s-6)(c-a^2)+2(a-1)^2\, . \label{fbfc}
\ee
All the models of this paper lie on the curve $f^{(s)}_b=0$, on which the term $h'\,\Box\,h'$ drops out of (\ref{lsplit}) and $f^{(s)}_c=(D+2s-6)(c-3a^2+2b)$. They are summarized in Table~\ref{tab:models}.

For $a\neq a^*_s$ the $r$-shift brings $(a,b)$ to $(1,1)$, and one obtains either the Fronsdal model, for $c=3a^2-2b$, or the partially broken Fronsdal (PBF) model \cite{pbf},
\be
\mathcal{L}_{PBF}^{(s)}(c) \equiv \mathcal{L}^{(s)}(1,1,c) = \mathcal{L}_F^{(s)} + \frac{3\,(c-1)}{4}\binom{s}{3}(\p\cdot h')^2\, , \label{lpbf}
\ee
with $c-1=(D+2s-6)^2\,(c-3a^2+2b)/[a(D+2s-4)-2]^2$. At $a=a^*_s$ the $r$-shift is not available, and the condition $f^{(s)}_b=0$ fixes $b$. The corresponding point is
\be
a^*_s=\frac{2}{D+2(s-2)}\, , \qquad b^*_s=\frac{D+2(s-1)}{[D+2(s-2)]^2}\, , \qquad c^*_s=3a^{*2}_s-2b^*_s=-\frac{2[D+2(s-4)]}{[D+2(s-2)]^2}\, . \label{svpoint}
\ee
At $c=c^*_s$ one finds the Skvortsov--Vasiliev model, $\mathcal{L}^{(s)}_{SV}\equiv\mathcal{L}^{(s)}(a^*_s,b^*_s,c^*_s)$ \cite{sv}, and for $c\neq c^*_s$ the partially broken SV (PBSV) model of \cite{dmmb},
\be
\mathcal{L}^{(s)}_{PBSV}(c)\equiv\mathcal{L}^{(s)}(a^*_s,b^*_s,c)=\mathcal{L}^{(s)}_{SV}+\frac{3\,(c-c^*_s)}{4}\binom{s}{3}(\p\cdot h')^2\, . \label{lpbsv}
\ee

\begin{table}[t]
	\centering
	\renewcommand{\arraystretch}{1.35}
	\begin{tabular}{@{}lcccc@{}}
		\toprule
		Model & $(a,b)$ & $c$ & Gauge symmetry & Restriction \\
		\midrule
		Fronsdal & $(1,1)$ & $1$ & $\delta h=\p\,\olamb$ & --- \\
		PBF & $(1,1)$ & $c\neq1$ & $\delta h=\p\,\lbdd$ & $\p\cdot\p\cdot\lbdd=0$ \\
		\midrule
		SV & $(a^*_s,b^*_s)$ & $c^*_s$ & $\delta h=\p\,\bar\Lambda^T+\eta\,\bar\sigma$ & --- \\
		PBSV & $(a^*_s,b^*_s)$ & $c\neq c^*_s$ & $\delta h=\p\,\bar\Lambda^T+\eta\,\bar\sigma^T$ & $\p\cdot\bar\sigma^T=0$ \\
		\bottomrule
	\end{tabular}
	\caption{Massless models on the curve $f^{(s)}_b=0$. All gauge parameters are traceless, $\bar\sigma$ has rank $s-2$, and the label $T$ denotes a transverse parameter. For $a\neq a^*_s$ the pair $(a,b)$ is brought to $(1,1)$ by the $r$-shift (\ref{rshift}), which is not available at $a=a^*_s$. In both lines the enhanced symmetry occurs at $c=3a^2-2b$. Since $\delta(\p\cdot h')$ is proportional to $\p\cdot\p\cdot\olamb$ in the first line and to $\p\cdot\bar\sigma$ in the second, the deformation $c\neq3a^2-2b$ restricts the gauge parameters by a traceless tensor of rank $s-3$.}\label{tab:models}
\end{table}

In both partially broken models the deformation multiplies $(\p\cdot h')^2$, where $\p\cdot h'$ is a tensor of rank $s-3$, so that only the sectors of spin $j\le s-3$ are affected. In the massless theories these sectors carry no propagating modes \cite{dmmb}, but in the massive case they must be made non-dynamical by the auxiliary fields: for $s=3$ only the spin-0 sector is involved, while for $s=4$ both the spin-1 and the spin-0 sectors are.

The reduced symmetry of the partially broken models weakens the Noether identity and allows for a larger class of sources. For the spin-3 PBF model they only have to satisfy \cite{pbf}
\be
\p_\mu T^{\mu\nu\alpha} = \eta^{\nu\alpha}\,\Omega + \p^\nu\p^\alpha J\, , \label{sources}
\ee
with arbitrary scalars $\Omega$ and $J$, while the Fronsdal theory requires $J=0$. The PBF and PBSV models propagate only massless spin-$s$ particles for any $s$ \cite{pbf,dmmb}, and their two-point amplitudes, computed in \cite{dmmb} for $s=3$ and $s=4$, coincide with the Fronsdal ones up to contact terms proportional to $1/(c-1)$ and $1/(c-c^*_s)$, respectively. For the spin-3 PBF model, for instance,
\be
\bar T\, G^{-1}\, T = \underbrace{\frac{2}{\Box}\left[\left(\widetilde{T}^{\mu\nu\lambda}\right)^2 - \frac{3}{D}\left(\widetilde{T}^{\mu}\right)^2\right]}_{\rm Fronsdal} + \frac{3}{(c-1)}\, J^2 \, , \label{amp3}
\ee
where $\widetilde{T}^{\mu\nu\lambda} = T^{\mu\nu\lambda} - \eta^{(\mu\nu}\p^{\lambda)}J/2$ satisfies the Fronsdal constraint. Here and below the sources are coupled through $h\cdot T$, and $\bar T\,G^{-1}\,T=-4W$, where $W$ is the action evaluated on the solution of the equations of motion in the presence of the sources; for $\mathcal{L}=\frac12\,h\,\Box\,h+h\cdot T$ this gives $\frac{2}{\Box}\,T\cdot T$, the normalization of the first term of (\ref{amp3}).

\section{Massive completions of the Fronsdal model}\label{sec:shdy}

Singh and Hagen (SH) \cite{sh,sh2} generalized the Fierz--Pauli description \cite{fp} of a massive spin-2 particle to arbitrary spin. For integer $s\ge3$, the conditions $(\Box-m^2)\phi=0$, $\p\cdot\phi=0$ and $\phi'=0$ on a rank-$s$ field cannot be obtained from a Lagrangian for $\phi$ alone; SH obtained them by introducing a tower of traceless auxiliary fields of rank $s-2,\dots,0$, which vanish on shell. In the massless limit the rank-$s$ and rank-$(s-2)$ fields combine into the double traceless field of the Fronsdal model \cite{fronsdal}.

For $s=3$ and $s=4$, a particularly transparent form of the SH model was derived by Aragone, Deser and Yang (ADY) \cite{ady} from the Kaluza--Klein reduction of the massless Fronsdal theory in $D+1$ dimensions, and it will be our reference throughout. Its kinetic term is the Fronsdal Lagrangian of a double traceless field $h$, whose trace $h'$ plays the role of the rank-$(s-2)$ auxiliary field of SH. The Noether identity of the Fronsdal Lagrangian states that the traceless part of the divergence of its equations of motion vanishes identically. Once a mass term $-\frac{m^2}{2}(h\cdot h+\alpha\,h'\cdot h')$ is added, this identity becomes a constraint which removes the spin-$(s-1)$ components, while $\alpha$ is fixed, as in the Fierz--Pauli case, by the absence of spin-$(s-2)$ modes, giving $\alpha=-3$ for $s=3$ and $\alpha=-6$ for $s=4$. The remaining modes, of spin $j\le s-3$, are removed by auxiliary fields of rank $s-3,\dots,0$, coupled to $h$ and among themselves through terms linear in $m$. For $s=3$ and $s=4$ we follow the notation of \cite{ady}, in which $a_r$ denotes an auxiliary field of rank $r$.

 For $s=3$ the ADY Lagrangian is
\be
\mathcal{L}_{ADY}^{(3)} = \mathcal{L}_F^{(3)} - \frac{m^2}{2}\left(h\cdot h - 3\, h'\cdot h'\right) + m\, a_0\, \p\cdot h' - \frac{\beta}{2}\, a_0\,\Box\, a_0 + \frac{\gamma}{2}\, m^2\, a_0^2 \, , \label{dy3}
\ee
where $\mathcal{L}_F^{(3)}$ is the Fronsdal Lagrangian (\ref{labc}) at $a=b=c=1$, and
\be
\beta = \frac{4}{3}\,\frac{D-1}{D-2}\, , \qquad \gamma = \frac{2D^2}{(D-2)^2}\, . \label{betagamma}
\ee
For $s=4$ it reads
\bea
\mathcal{L}_{ADY}^{(4)} &=& \mathcal{L}_F^{(4)} - \frac{m^2}{2}\left(h\cdot h - 6\, h'\cdot h'\right) + 12\, m\, a_1\cdot\p\cdot h' \nn\\ 
&+& \frac32\Big[ -\beta_4\, a_1\cdot\Box\, a_1 + \gamma_4\, (\p\cdot a_1)^2 - \sigma_4\, a_0\,\Box\, a_0 \nn\\
&&\qquad+\, m^2\left(\tau_4\, a_1\cdot a_1 + \rho_4\, a_0^2\right) - 2\, m\, (\p\cdot a_1)\, a_0 \Big] \, , \label{dy4}
\eea
with $a_1$ and $a_0$ normalized as in (\ref{dy4}) and\footnote{The coefficients (\ref{dy4coef}) and the coupling $12m\,a_1\cdot\p\cdot h'$ are those of eqs.~(4.4)--(4.6) of \cite{ady}, but the coefficient $-2m$ of the mixing term $(\p\cdot a_1)\,a_0$ in (\ref{dy4}) differs from the value $-3m$ printed in eq.~(4.5) of \cite{ady}. The value $-2m$ is the one required by the criterion (\ref{criterion}) below: with $-3m$, the determinant of the spin-0 sector is not constant. It is also the value obtained in Section~\ref{sec:ansatz4} by eliminating an algebraic field from the minimal description of the spin-4 PBF model, and it was found independently in \cite{pbfkk}.}
\be
\begin{aligned}
	&\beta_4 = \frac{16(D+1)}{D}\, , \qquad \gamma_4 = \frac{8(D-2)}{D}\, , \qquad \tau_4 = \frac{24(D+2)^2}{D^2}\, ,\\
	&\sigma_4 = \frac{1}{8}\,\frac{(D-1)\,D^2}{(D^2-4)(D+3)}\, , \qquad \rho_4 = \frac{1}{4}\,\frac{D^3(D+1)}{(D-2)^2(D+2)(D+3)}\, .
\end{aligned} \label{dy4coef}
\ee

The coefficients of (\ref{dy3}) and (\ref{dy4}) can be recovered from a simple criterion, which we will use throughout. Since the theory is Lorentz invariant, its equations of motion do not mix components of different spin $j$. The spin-$j$ variables of each field are obtained by taking divergences and traces and projecting onto transverse traceless structures; for a rank-3 field they are $\p\cdot\p\cdot\p\cdot h$ and $\p\cdot h'$ for $j=0$, $(\p\cdot\p\cdot h)^T$ and $h'^{\,T}$ for $j=1$, $(\p\cdot h)^{TT}$ for $j=2$ and $h^{TT}$ for $j=3$. Applying the same operations to the equations of motion gives, in each sector, a square matrix $M^{(j)}(\Box)$ whose entries are polynomials in $\Box$ and $m$, and nontrivial plane-wave solutions in the spin-$j$ sector can only exist at the zeros of $\det M^{(j)}$. In the absence of gauge symmetries, or after gauge fixing, a single massive spin-$s$ particle is therefore ensured by
\be
\det M^{(s)} \propto (\Box-m^2) \, , \qquad \det M^{(j)} = \mathrm{const}\neq 0 \, , \quad j<s\, , \label{criterion}
\ee
since a constant determinant makes $M^{(j)}$ invertible by a local operator. This is the requirement of \cite{ady} that all lower-spin excitations be removed. For $s=3$, for instance, the spin-2 sector is non-dynamical because of the Noether identity of the Fronsdal kinetic term, the spin-1 sector because of $\alpha=-3$, and $\beta$ and $\gamma$ follow from the spin-0 sector. In general, (\ref{criterion}) is solved from the highest power of $\Box$ downwards. Field redefinitions with a local inverse, such as invertible algebraic mixings of fields of the same type or the shift of a field by derivatives of another one, multiply $\det M^{(j)}$ by a nonzero constant, so that the solutions are determined up to such redefinitions.

The PBF model has the same particle content as Fronsdal's, but a smaller gauge symmetry, with the restricted parameter $\lbdd$ (Table~\ref{tab:models}), and hence a weaker Noether identity. Since the deformation only affects the sectors of spin $j\le s-3$ (Section~\ref{sec:pbf}), the sectors of spin $s$, $s-1$ and $s-2$ are those of the ADY model, and they fix the same mass term. Nothing guarantees, however, that the SH auxiliary content is still sufficient for $c\neq1$. What, then, is the massive completion of the PBF model? In the next sections we look for a massive spin-$s$ theory whose kinetic sector is $\mathcal{L}_{PBF}^{(s)}(c)$, with $c\neq1$, and which reduces to the ADY model at $c=1$. We add a mass term with an arbitrary trace coefficient and the most general local auxiliary sector with at most two derivatives, and let the criterion (\ref{criterion}) decide which auxiliary fields are needed, enlarging the auxiliary content only when smaller sets are proven to fail. The absence of ghosts follows, for all the solutions we find, from their equivalence with the ADY model, and hence with the SH model, whose unitarity in arbitrary dimension was established in \cite{SchimidtElias}. All computations were carried out with xAct/xTras \cite{teake}.
\section{Massive PBF models: spin 3 and spin 4}\label{sec:spin3}
	
	\subsection{Spin-3}\label{sec:twoscalars}
	
	For $s=3$ the PBF Lagrangian (\ref{lpbf}) reads
	\be
	\mathcal{L}_{PBF}^{(3)}(c) = \frac 12 h\, \Box \, h + \frac 32 (\p\cdot h)^2 + 3\, h' \, \p\cdot\p\cdot h - \frac 32\, h'\, \Box \, h' + \frac 34\, c \, (\p\cdot h')^2 \, , \label{lpbf3}
	\ee
	and we add to it the mass term
	\be
	\mathcal{L}_m = -\frac{m^2}{2}\left( h_{\mu\nu\alpha}h^{\mu\nu\alpha} + q_1 \, h'_\mu h'^\mu\right) \, , \label{lm3}
	\ee
	where $q_1$ plays the role of the coefficient $\alpha$ of the ADY construction. Since $c$ only affects the spin-0 sector, we look for scalar auxiliary fields. Since a single scalar is the only auxiliary content with fewer components than two scalars, the minimality of the solution found below holds in the number of auxiliary components, independently of their tensor type.\footnote{Auxiliary vectors can also be used. With a vector $A$ and a scalar $\pi$ we find the solution
		\[
		\mathcal{L}^{(3)}_{PBF}(c)-\frac{m^2}{2}\big(h\cdot h-3\,h'\cdot h'\big)-\frac{D}{12(D+1)}F_{\mu\nu}F^{\mu\nu}+\zeta\, m^2\Big(A+\frac{\p\pi}{2\zeta m}\Big)^2+m^2A\cdot h'+\frac{3(D-2)^2}{16(D-1)^2(c-1)}\,m^2\pi^2\, ,
		\]
		with $\zeta=\frac{2(D-1)}{3(D-2)}$, whose spin-1 and spin-0 determinants are constant for every $c\neq1$. The redefinition $A\to A-\p\pi/(2\zeta m)$ makes $\pi$ algebraic, and its elimination gives back the Fronsdal kinetic term together with the massive spin-3 model with an algebraically coupled auxiliary vector of \cite{bvr}, given for arbitrary $D$ in Appendix~B of \cite{ady}. As discussed there, this model is physically equivalent to the minimal one with a scalar auxiliary field, although the two are not related by a field redefinition at the level of the covariant action. The mechanism by which $c$ disappears is therefore the same as for the scalar solution. An analogous analysis for $s=4$ with an auxiliary field of rank 2 involves a much larger set of possible terms and has not been completed.} A scalar built linearly from the rank-3 field $h$ contains an odd number of derivatives, so, with at most two derivatives in the Lagrangian, the only coupling between a scalar $\vf$ and $h$ is, up to total derivatives, $m\,\vf\,\p\cdot h'$.
	
	With two scalars $\vf$ and $\chi$, a rotation in field space allows us to couple only $\vf$ to $h$, and the shift $\chi\to\chi+\epsilon\,\vf$ removes the kinetic mixing $\vf\,\Box\,\chi$.\footnote{The shift requires $\chi$ to have a kinetic term of its own. In the degenerate case in which $\chi$ only appears through a kinetic mixing $k\,\vf\,\Box\,\chi$, the coefficient of the highest power of $\Box$ in the spin-0 determinant is proportional to $(D-1)(c-1)k^2$, so that $k=0$ for $c\neq1$ and we are back to (\ref{l3ansatz}).} The most general Lagrangian is then
	\bea
	\mathcal{L} &=& \mathcal{L}_{PBF}^{(3)}(c) + \mathcal{L}_m + q_2\, m\, \vf\, \p\cdot h' + \frac{q_3}{2}\, \vf\,\Box\,\vf - \frac{q_4}{2}\, m^2 \vf^2 \nn\\
	&+& \frac{q_5}{2}\, \chi\,\Box\,\chi - \frac{q_6}{2}\, m^2 \chi^2 - q_7\, m^2\, \vf\,\chi \, . \label{l3ansatz}
	\eea
	The case of a single scalar is obtained by removing $\chi$, and will be discussed at the end of this subsection.
	
The auxiliary scalars only carry spin 0. Their contribution to the equation of motion of $h$, $E_h$,\footnote{Throughout, $E_h$, $E_\vf$ and $E_\chi$ denote the Euler--Lagrange derivatives $\delta S/\delta h$, $\delta S/\delta\vf$ and $\delta S/\delta\chi$. Divergences, traces and projections of $E_h$ are always taken on its free indices.} is proportional to $\eta_{(\mu\nu}\p_{\rho)}\vf$ and is removed by the transverse projections, so that the sectors of spin 3, 2 and 1 involve $h$ alone. The spin-3 sector is $E_h^{TT}=(\Box-m^2)\,h^{TT}$, so that $\det M^{(3)}=\Box-m^2$. In the spin-2 sector, the Noether identity of the PBF symmetry $\delta h=\p\,\lbdd$ (Table~\ref{tab:models}) states that $(\p\cdot E_h)^{TT}$ vanishes identically for the kinetic part of $E_h$, as for the Fronsdal model, since the restriction $\p\cdot\p\cdot\lbdd=0$ does not involve the transverse parameters. With the mass term, $(\p\cdot E_h)^{TT}=0$ becomes the algebraic constraint $m^2(\p\cdot h)^{TT}=0$, and $\det M^{(2)}=-m^2$, irrespective of $q_1$ and $c$.

The spin-1 sector is where the mass term is tuned. Taking $(\p\cdot\p\cdot E_h)^T$ and $E_h'^{\,T}$ as rows, and $h'^{\,T}$ and $(\p\cdot\p\cdot h)^T$ as columns, we find
\be
M^{(1)}=\begin{pmatrix} -\frac{q_1}{3}\,m^2\Box & -m^2 \\[2pt] -\frac13\left[3+(D+2)q_1\right]m^2 - D\,\Box & D\end{pmatrix},
\ee
whose determinant is
\be
\det M^{(1)} = -\frac{D}{3}\,(3+q_1)\, m^2\, \Box-\frac13\left[3+(D+2)q_1\right]m^4\, .
\ee
The term linear in $\Box$ vanishes only for $q_1=-3$, the ADY value, and then $\det M^{(1)}=(D+1)\,m^4\neq0$. The parameter $c$ plays no role: the deformation contributes to $E_h$ only through $\eta_{(\mu\nu}\p_{\rho)}(\p\cdot h')$, whose trace and double divergence are gradients of a scalar. From now on we set $q_1=-3$.

The spin-0 sector is the only one sensitive to $c$, and it is where the auxiliary fields are needed. Taking $E_\vf$, $E_\chi$, $\p\cdot\p\cdot\p\cdot E_h$ and $\p\cdot E_h'$ as rows, and $\vf$, $\chi$, $\p\cdot h'$ and $\p\cdot\p\cdot\p\cdot h$ as columns, we find
\be
M^{(0)}=\begin{pmatrix}
	q_3\Box-q_4m^2 & -q_7m^2 & q_2\, m & 0\\[2pt]
	-q_7m^2 & q_5\Box-q_6m^2 & 0 & 0\\[2pt]
	-q_2\, m\,\Box^2 & 0 & -\frac32\,\Box\left(c\,\Box-2m^2\right) & \Box-m^2\\[2pt]
	-\frac{D+2}{3}\,q_2\, m\,\Box & 0 & (D+1)\,m^2 - \frac12\big[2(c-1)+D(c+2)\big]\Box & D
\end{pmatrix}. \label{M0}
\ee
Without auxiliary fields only the $2\times2$ block of $h$ is left, whose determinant is $-\mathcal{A}$, with
\be
\mathcal{A} = (D-1)(c-1)\,\Box^2 + \left[\tfrac12(D+2)\,c - D\right] m^2\Box - (D+1)\, m^4 \, . \label{Ablock}
\ee
It depends on $\Box$ even at $c=1$: without auxiliary fields, the spin-0 sector of a massive spin-3 field always propagates.

The determinant of (\ref{M0}) is a polynomial of degree four in $\Box$, and (\ref{criterion}) requires the coefficients of $\Box^4,\dots,\Box$ to vanish and the constant term to be nonzero. We impose these conditions in cascade, starting from the highest power and substituting each solution before computing the next coefficient; at each step the coefficient is linear in one of the parameters. The coefficient of $\Box^4$ is $-(D-1)(c-1)\,q_3\,q_5$, so for $c\neq1$ either $q_3$ or $q_5$ must vanish. The branch $q_5=0$ does not work: if $q_6\neq0$, the field $\chi$ is algebraic and its elimination only shifts $q_4\to q_4-q_7^2/q_6$, which reduces the problem to the single-scalar case discussed below, while for $q_5=q_6=0$ the determinant is $q_7^2m^4\mathcal{A}$, which is not constant. Hence $\chi$ must be dynamical, and
\be
q_3 = 0 \, , \label{sol1}
\ee
that is, the scalar which couples to $h$ is non-dynamical. The coefficient of $\Box^3$ then becomes $\frac13(D-1)\,q_5\left[2q_2^2+3(c-1)q_4\right]m^2$, and its vanishing fixes the mass term of $\vf$,
\be
q_4 = -\frac{2\,q_2^2}{3(c-1)}\, . \label{sol2}
\ee
The coefficient of $\Box^2$ is then $\big[(D-2)\,q_2^2\,q_5+3(D-1)(c-1)^2q_7^2\big]m^4/[3(c-1)]$, which vanishes for\footnote{The coefficients of $\Box^4$ and $\Box^3$ do not depend on $q_1$, which first enters at order $\Box^2$. For arbitrary $q_1$ this step gives $q_5=9(D-1)(c-1)^2q_7^2/\{[9D-6+4(D-1)q_1]\,q_2^2\}$, which reduces to (\ref{sol3}) for $q_1=-3$.}
\be
q_5 = -\frac{3(D-1)(c-1)^2}{D-2}\,\frac{q_7^2}{q_2^2}\, . \label{sol3}
\ee
For $D>2$ the kinetic term of $\chi$ has the same sign as that of the ADY scalar $a_0$ in (\ref{dy3}). The coefficient of $\Box$ is then $-\big[2(D-2)^2q_2^2\,q_6+3(c-1)(3D^2c-2D^2-4D+4)\,q_7^2\big]m^6/[6(D-2)(c-1)]$, which vanishes for
\be
q_6 = -\frac{3(c-1)\left(3D^2c-2D^2-4D+4\right)}{2(D-2)^2}\,\frac{q_7^2}{q_2^2}\, . \label{sol4}
\ee
Only the constant term, $(D+1)(q_4q_6-q_7^2)\,m^8$, survives, and
\be
\det M^{(0)} = \frac{3D^2(D+1)(c-1)}{(D-2)^2}\, q_7^2\, m^8 \, , \label{detfinal}
\ee
which is nonzero for every $c\neq1$ and $q_7\neq0$.

The two parameters left free, $q_2$ and $q_7$, are the normalizations of $\vf$ and $\chi$, and neither can vanish: for $q_2=0$ the field $\vf$ decouples from $h$, and for $q_7=0$ the determinant (\ref{detfinal}) vanishes.

The same cascade settles the case of a single auxiliary scalar. Deleting the row and the column of $\chi$ in (\ref{M0}), the determinant is of degree three in $\Box$. Its coefficients of $\Box^3$ and $\Box^2$ give again $q_3=0$ and $q_4=-2q_2^2/[3(c-1)]$, and the coefficient of $\Box$ is then $(D-2)\,q_2^2\,m^4/[3(c-1)]$, which cannot vanish for $D\geq3$ and $q_2\neq0$. Hence no theory with a single auxiliary scalar describes a massive spin-3 particle when $c\neq1$, and two scalars are the minimal content.

With the normalizations $q_2=c-1$ and $q_7=2/3$, one has $q_5=-\beta$, with $\beta$ given in (\ref{betagamma}), and (\ref{l3ansatz}) becomes
\bea
\mathcal{L}^{(3)} &=& \mathcal{L}_{PBF}^{(3)}(c) - \frac{m^2}{2}\left(h\cdot h - 3\, h'\cdot h'\right) + (c-1)\, m\, \vf\, \p\cdot h' + \frac{c-1}{3}\, m^2 \vf^2 - \frac23\, m^2\, \vf\,\chi \nn\\
&-& \frac{\beta}{2}\, \chi\,\Box\,\chi + \left[\frac{\gamma}{2} + \frac{1}{3(c-1)}\right] m^2\chi^2 \, . \label{l3final}
\eea
The field $\vf$ is algebraic, with $\vf=\chi/(c-1)-3\,\p\cdot h'/(2m)$ on shell. Its elimination produces $-\frac34(c-1)(\p\cdot h')^2$, which cancels the deformation of $\mathcal{L}_{PBF}^{(3)}(c)$, generates the coupling $m\,\chi\,\p\cdot h'$, and cancels the singular part of the mass term of $\chi$. The result is exactly the ADY Lagrangian (\ref{dy3}), with $\chi=a_0$, for every value of $c$. This does not contradict the single-scalar result, since after the elimination of $\vf$ the kinetic term of $h$ is Fronsdal's: within the class (\ref{l3ansatz}), one scalar must be algebraic and absorb the deformation, while the other plays the role of the ADY scalar.

The coefficients of (\ref{l3final}) are singular at $c=1$, because the basis in which only $\vf$ couples to $h$ is adapted to $c\neq1$. The redefinition $\vf=\sigma+\chi/(c-1)$ removes both the mixing and the singular part of the mass term of $\chi$, and gives
\be
\mathcal{L}^{(3)} = \mathcal{L}_{ADY}^{(3)}\Big|_{\mathcal{L}_F\to\mathcal{L}_{PBF}(c)} + (c-1)\left[ m\,\sigma\,\p\cdot h' + \frac13\, m^2 \sigma^2 \right] , \label{l3regular}
\ee
whose coefficients are all regular. At $c=1$ the ADY model is recovered exactly, and for $c\neq1$ the elimination of $\sigma=-3\,\p\cdot h'/(2m)$ cancels the deformation of the kinetic term.

	\subsection{Spin-4}\label{sec:ansatz4}
	
	For $s=4$ the deformation in (\ref{lpbf}) is $3(c-1)(\p\cdot h')^2$. Since $\p\cdot h'$ is a vector, it affects both the spin-1 and the spin-0 sectors, while the sectors of spin 4, 3 and 2 coincide with those of the ADY model. The trace coefficient of the mass term is therefore fixed as in \cite{ady}, and we take from the start
	\be
	\mathcal{L}_m=-\frac{m^2}{2}\left(h\cdot h-6\,h'\cdot h'\right) . \label{lm4}
	\ee
	
	The spin-3 result suggests the auxiliary content: the ADY fields, here a vector and a scalar, supplemented by one field of rank $s-3$, here a second vector. We therefore consider two vectors, $\mathcal{A}$ and $B$, and one scalar, $\chi$, without assuming in advance that any of them is algebraic.\footnote{We have also analysed a larger ansatz, with two scalars and a Maxwell kinetic term for $B$. It leads to the same theory: one of the two scalars then reproduces the longitudinal part of the kinetic term of $B$, which is included from the start in (\ref{l4ansatz}).} With at most two derivatives, the only vector and scalar built linearly from $h$ are $\p\cdot h'$ and $\p\cdot\p\cdot h'$, and a linear redefinition of the vectors allows only $\mathcal{A}$ to couple to $\p\cdot h'$. Up to total derivatives, the most general Lagrangian is
	\bea
	\mathcal{L} &=& \mathcal{L}_{PBF}^{(4)}(c) + \mathcal{L}_m + 12\, m\, \mathcal{A}\cdot\p\cdot h' + g\,\chi\,\p\cdot\p\cdot h' \nn\\
	&-& u_1\,(\p\cdot\mathcal{A})^2 + u_2\,\mathcal{A}\cdot\Box\mathcal{A} + u_3\, m^2\, \mathcal{A}\cdot\mathcal{A} \nn\\
	&+& w_1\, F_{\mu\nu}(B)F^{\mu\nu}(B) - v_1\,(\p\cdot B)^2 + w_3\, m^2\, B\cdot B \nn\\
	&+& w_2\, F_{\mu\nu}(\mathcal{A})F^{\mu\nu}(B) - v_2\,(\p\cdot\mathcal{A})(\p\cdot B) + w_4\, m^2\, \mathcal{A}\cdot B \nn\\
	&+& m\left(z_1\,\mathcal{A}+z_2\,B\right)\cdot\p\chi + s_0\,\chi\,\Box\,\chi + r_0\, m^2\,\chi^2 , \label{l4ansatz}
	\eea
	where $F_{\mu\nu}(X)=\p_\mu X_\nu-\p_\nu X_\mu$. The normalization of $\mathcal{A}$ is fixed by its coupling to $h$, while $w_1$ and $z_2$ are the normalizations of $B$ and $\chi$.
	
	The contributions of the auxiliary fields and of the deformation to the equation of motion of $h$\footnote{Since $h$ is double traceless, its Euler--Lagrange derivative is defined up to terms proportional to $\eta_{(\mu\nu}\eta_{\rho\sigma)}$. Throughout, $E_h$ denotes the double traceless projection of $\delta S/\delta h$, so that $E_h''=0$ identically.} are removed by the transverse traceless projections, and the sectors of spin 4, 3 and 2 involve $h$ alone. As for $s=3$, $\det M^{(4)}=\Box-m^2$, and the Noether identity of the PBF symmetry, now with $\lbdd$ of rank 3, turns $(\p\cdot E_h)^{TT}=0$ into the algebraic constraint $m^2(\p\cdot h)^{TT}=0$, so that $\det M^{(3)}=-m^2$. In the spin-2 sector, taking $E_h'^{\,TT}$ and $(\p\cdot\p\cdot E_h)^{TT}$ as rows, and $h'^{\,TT}$ and $(\p\cdot\p\cdot h)^{TT}$ as columns, we find
	\be
	M^{(2)}=\begin{pmatrix} (D+3)\,m^2-(D+2)\,\Box & D+2 \\[2pt] m^2\,\Box & -m^2\end{pmatrix},
	\ee
	with $\det M^{(2)}=-(D+3)\,m^4$. For a generic trace coefficient $\alpha$ the determinant contains the term $(D+2)\left(1+\frac{\alpha}{6}\right)m^2\Box$, which vanishes only for $\alpha=-6$, as in (\ref{lm4}).
	
	In the spin-1 sector $u_1$, $v_1$, $v_2$ and the couplings of $\chi$ drop out, since they only multiply gradients. Taking the transverse parts of $E_{\mathcal{A}}$, $E_B$, $\p\cdot\p\cdot\p\cdot E_h$ and $\p\cdot E_h'$ as rows, and $\mathcal{A}^T$, $B^T$, $(\p\cdot h')^T$ and $(\p\cdot\p\cdot\p\cdot h)^T$ as columns, we find
	\be
	M^{(1)}=\begin{pmatrix}
		2\left(u_3m^2+u_2\Box\right) & w_4m^2-2w_2\Box & 12\,m & 0\\[2pt]
		w_4m^2-2w_2\Box & 2w_3m^2-4w_1\Box & 0 & 0\\[2pt]
		-3\,m\,\Box^2 & 0 & 3\,\Box\left(m^2-\frac{c}{2}\,\Box\right) & \Box-m^2\\[2pt]
		-(D+4)\,m\,\Box & 0 & (D+3)\,m^2-\frac12\left[D(c+2)+4c+2\right]\Box & D+2
	\end{pmatrix}. \label{M1s4}
	\ee
	Its determinant has degree four in $\Box$, and we solve (\ref{criterion}) in cascade, as for $s=3$. The coefficient of $\Box^4$ is $4(D+1)(c-1)\left(2u_2w_1+w_2^2\right)$: it is the deformation which raises the degree of the determinant. Since $B$ does not couple to $h$, the redefinition $B\to B-\frac{w_2}{2w_1}\mathcal{A}$ preserves the form of (\ref{l4ansatz}) and sets $w_2=0$, and then $u_2=0$: the transverse part of $\mathcal{A}$ is non-dynamical, as the scalar $\vf$ for $s=3$. The coefficient of $\Box^3$ is then $8(D+1)\,w_1\left[(c-1)\,u_3-12\right]m^2$, the coefficient of $\Box^2$ becomes $\left[(D+1)(c-1)\,w_4^2-48D\,w_1/(c-1)\right]m^4$, and the coefficient of $\Box$ is linear in $w_3$. They give, one at a time,
	\be
	u_3=\frac{12}{c-1}\, , \qquad w_4^2=\frac{48D\,w_1}{(D+1)(c-1)^2}\, , \qquad w_3=\frac{w_1}{D(D+1)}\left[3(D+2)^2+\frac{D^2}{c-1}\right] , \label{sol4vec}
	\ee
	where the two roots for $w_4$ are related by $B\to-B$, and reality requires $w_1>0$. Only the constant term survives,
	\be
	\det M^{(1)}=\frac{144\,(D+2)^2(D+3)}{D(D+1)(c-1)}\,w_1\,m^8\, ,
	\ee
	which is nonzero for every $c\neq1$.\footnote{The second vector is necessary. Scalars do not enter the spin-1 sector, and with a single vector $\mathcal{A}$ the spin-1 matrix is obtained from (\ref{M1s4}) by deleting the row and the column of $B$. The coefficients of $\Box^3$ and $\Box^2$ of its determinant give $u_2=0$ and $u_3=12/(c-1)$, after which the coefficient of $\Box$ is $12D\,m^4/(c-1)\neq0$. A single vector, with any number of scalars, is therefore never sufficient for $c\neq1$.}
	
	The spin-0 sector has five variables. Taking $E_\chi$, $\p\cdot E_{\mathcal{A}}$, $\p\cdot E_B$, $\p\cdot\p\cdot E_h'$ and $\p\cdot\p\cdot\p\cdot\p\cdot E_h$ as rows, and $\p\cdot\p\cdot h'$, $\p\cdot\p\cdot\p\cdot\p\cdot h$, $\p\cdot\mathcal{A}$, $\p\cdot B$ and $\chi$ as columns, we find, with $u_2=w_2=0$,
	\be
	M^{(0)}=\begin{pmatrix}
		g & 0 & -z_1\,m & -z_2\,m & 2(r_0m^2+s_0\Box)\\[2pt]
		12\,m & 0 & 2(u_3m^2+u_1\Box) & w_4m^2+v_2\Box & z_1\,m\,\Box\\[2pt]
		0 & 0 & w_4m^2+v_2\Box & 2(w_3m^2+v_1\Box) & z_2\,m\,\Box\\[2pt]
		e_1 & D+2 & -\frac{2(D+4)(D-1)}{D}\,m\,\Box & 0 & \frac{(D+4)(D-1)}{6D}\,g\,\Box^2\\[2pt]
		e_2 & 3\Box-m^2 & -\frac{12(D-1)}{D}\,m\,\Box^2 & 0 & \frac{D-1}{D}\,g\,\Box^3
	\end{pmatrix}, \label{M0s4}
	\ee
	where
	\be
	e_1=(D+3)\,m^2-\frac{D^2(c+1)+D(3c+1)+4(1-c)}{D}\,\Box\, ,\qquad e_2=\frac{6\,\Box}{D}\Big[D\,m^2-\big((D-1)c+1\big)\Box\Big] .
	\ee
	With (\ref{sol4vec}), the determinant has degree five in $\Box$, and the coefficient of $\Box^5$ is
	\be
	\frac{D-1}{2}\left(4u_1v_1-v_2^2\right)\left[g^2+12(c-1)\,s_0\right] .
	\ee
	The first factor is, up to normalization, the determinant of the longitudinal kinetic matrix of the two vectors, and the second has the same structure as in the spin-3 case. We take the branch in which the longitudinal kinetic matrix is degenerate,
	\be
	u_1=v_2=0\, , \label{u1zero}
	\ee
	so that $\mathcal{A}$ is purely algebraic and the only longitudinal kinetic term left is that of $B$, as for the ADY vector $a_1$. The coefficient of $\Box^4$ is then proportional to $v_1\left[2g+(c-1)\,z_1\right]^2 m^2$. Since $\mathcal{A}$ is algebraic, the redefinition $\mathcal{A}\to\mathcal{A}+\frac{\lambda}{m}\,\p\chi$ preserves the form of (\ref{l4ansatz}), shifting $g\to g-12\lambda$ and $z_1\to z_1+\frac{24\lambda}{c-1}$, so that $2g+(c-1)z_1$ is invariant. For $v_1\neq0$ it must vanish, and the choice $\lambda=g/12$ sets
	\be
	g=z_1=0\, :
	\ee
	the direct coupling of the scalar to $h$ is removable, and $\chi$ only couples to $B$, as $a_0$ does in the ADY model. The coefficient of $\Box^3$ is then proportional to $s_0\left[v_1+3D\,w_1/(D+1)\right]m^4$, and the coefficients of $\Box^2$ and $\Box$ are linear in $s_0$ and $r_0$, respectively. They give
	\bea
	&&v_1=-\frac{3D}{D+1}\,w_1\, ,\qquad s_0=-\frac{D(D^2-1)}{4(D^2-4)(D+3)}\,\frac{z_2^2}{w_1}\, ,\nn\\[4pt]
	&&r_0=\frac{D^2(D+1)^2}{2(D-2)^2(D+2)(D+3)}\,\frac{z_2^2}{w_1}\, . \label{sol4sc}
	\eea
	The alternative $s_0=0$ has no solution: the coefficient of $\Box^3$ then vanishes for every $v_1$, but the conditions from the coefficients of $\Box^2$ and $\Box$ force $r_0=0$, for which the coefficient of $\Box^2$ is a nonzero constant. For the same reason $z_2\neq0$. With (\ref{sol4sc}),
	\be
	\det M^{(0)}=-\frac{144\,D(D+1)(D+2)}{(D-2)^2(c-1)}\,z_2^2\,m^{10}\, ,
	\ee
	which is nonzero for every $c\neq1$. The theory therefore describes a single massive spin-4 particle for every $c\neq1$. Note that (\ref{sol4sc}) does not depend on $c$: the deformation only enters through the spin-1 coefficients (\ref{sol4vec}).
	
	With the normalizations $w_1=12(D+1)/D$ and $z_2=3$, and the identifications $B=-a_1$ and $\chi=-a_0$, one finds $u_3=12/(c-1)$, $w_4=24/(c-1)$ and $w_3=\frac32\,\tau_4+\frac{12}{c-1}$, while the remaining coefficients are precisely those of (\ref{dy4coef}). The terms singular at $c=1$ combine into a perfect square, and the final Lagrangian reads
	\bea
	\mathcal{L}^{(4)} &=& \mathcal{L}_{PBF}^{(4)}(c) - \frac{m^2}{2}\left(h\cdot h - 6\, h'\cdot h'\right) + 12\, m\, \mathcal{A}\cdot\p\cdot h' + \frac{12}{c-1}\, m^2\left(\mathcal{A}-a_1\right)^2 \nn\\
	&+& \frac32\Big[ -\beta_4\, a_1\cdot\Box\, a_1 + \gamma_4\, (\p\cdot a_1)^2 - \sigma_4\, a_0\,\Box\, a_0 \nn\\
	&&\qquad+\, m^2\left(\tau_4\, a_1\cdot a_1 + \rho_4\, a_0^2\right) - 2\, m\, (\p\cdot a_1)\, a_0 \Big] . \label{l4final}
	\eea
	The vector $\mathcal{A}$ is algebraic, with $\mathcal{A}=a_1-\frac{c-1}{2m}\,\p\cdot h'$ on shell. Its elimination produces $-3(c-1)(\p\cdot h')^2$, which cancels the deformation, and gives exactly the ADY Lagrangian (\ref{dy4}) for every $c$.
	
	As for $s=3$, the singularity at $c=1$ is an artifact of the basis. The redefinition $\mathcal{A}=a_1+\frac{c-1}{12}\,\sigma$ gives
	\be
	\mathcal{L}^{(4)} = \mathcal{L}_{ADY}^{(4)}\Big|_{\mathcal{L}_F\to\mathcal{L}_{PBF}(c)} + (c-1)\left[ m\,\sigma\cdot\p\cdot h' + \frac{1}{12}\, m^2\, \sigma\cdot\sigma \right] , \label{l4regular}
	\ee
	whose coefficients are all regular. This is the structure found for spin 3, with the algebraic scalar replaced by an algebraic vector: the auxiliary content of the massive spin-4 PBF model is the ADY content supplemented by one algebraic vector, which absorbs the deformation.

	\section{Massive PBF models: arbitrary spin}\label{sec:spins}\label{sec:anys}
	
	The final Lagrangians (\ref{l3regular}) and (\ref{l4regular}) only use two properties of the ADY model: that $h$ enters it through the undeformed Fronsdal kinetic term, and that the deformation of the PBF model is a quadratic form in $\p\cdot h'$. The detailed form of the auxiliary sector plays no role, and the construction extends to every $s\geq3$.
	
	For arbitrary $s$, the dimensional reduction of \cite{ady} gives a Lagrangian of the form
	\be
	\mathcal{L}_{ADY}^{(s)}=\mathcal{L}_F^{(s)}-\frac{m^2}{2}\Big(h\cdot h-\frac{s(s-1)}{2}\,h'\cdot h'\Big)+g_s\,m\,a_{s-3}\cdot\p\cdot h'+\mathcal{L}_{aux}(a_{s-3},a_{s-4})\, , \label{ladys}
	\ee
	where $g_s$ depends on the normalization of $a_{s-3}$ ($g_3=1$ and $g_4=12$ in (\ref{dy3}) and (\ref{dy4})). The auxiliary fields $a_{s-3}$ and $a_{s-4}$ are symmetric tensors of ranks $s-3$ and $s-4$, which are not traceless; decomposed into traceless parts, they contain one traceless tensor of each rank $s-3,\dots,0$, which, together with the trace $h'$, are the auxiliary fields of \cite{sh}. The explicit form of $\mathcal{L}_{aux}$ will not be needed. All we use is that (\ref{ladys}) describes a single massive spin-$s$ particle and that $h$ enters it through $\mathcal{L}_F^{(s)}$. The gauge invariant formulation of \cite{zinoviev} has the same property and could be used equally well.
	
	Let then $\mathcal{L}_{ADY}^{(s)}=\mathcal{L}_F^{(s)}+R[h,a]$, where $R$ collects the mass terms, the coupling and $\mathcal{L}_{aux}$, and introduce a symmetric traceless tensor $\sigma$ of rank $s-3$. We consider
	\be
	\mathcal{L}^{(s)}=\mathcal{L}_{PBF}^{(s)}(c)+R[h,a]+(c-1)\Big[m\,\sigma\cdot\p\cdot h'+\kappa\,m^2\,\sigma\cdot\sigma\Big] ,\qquad \kappa=\frac{2\,(s-3)!}{s!}\, , \label{lsgen}
	\ee
	that is, $\mathcal{L}_{ADY}^{(s)}\big|_{\mathcal{L}_F\to\mathcal{L}_{PBF}(c)}$ plus an algebraic field coupled to $\p\cdot h'$, which is traceless as a consequence of $h''=0$. For $s=3$ and $s=4$, (\ref{lsgen}) is (\ref{l3regular}) and (\ref{l4regular}). The equation of motion of $\sigma$ gives $\sigma=-\p\cdot h'/(2\kappa m)$, and substituting back produces $-\frac{c-1}{4\kappa}(\p\cdot h')^2=-\frac34(c-1)\binom{s}{3}(\p\cdot h')^2$, which is exactly minus the deformation in (\ref{lpbf}). This requirement fixes $\kappa$, and reproduces the values $\frac13$ and $\frac1{12}$ found for $s=3$ and $s=4$. Equivalently, since $\mathcal{L}_{PBF}^{(s)}(c)=\mathcal{L}_F^{(s)}+\frac{c-1}{4\kappa}(\p\cdot h')^2$,
	\be
	\mathcal{L}^{(s)} = \mathcal{L}^{(s)}_{ADY} + (c-1)\,\kappa\, m^2\left(\sigma+\frac{1}{2\kappa m}\,\p\cdot h'\right)^2 . \label{square}
	\ee
	After the elimination of $\sigma$, (\ref{lsgen}) reduces to $\mathcal{L}_{ADY}^{(s)}$ for every value of $c$, and at $c=1$ the field $\sigma$ decouples. Hence, for every $s\geq3$, every dimension $D$ and every $c$, the PBF model admits a massive completion which describes a single massive spin-$s$ particle, with the ADY auxiliary content supplemented by one algebraic traceless field of rank $s-3$.
	
	Two remarks clarify the scope of this result. First, the rank of $\sigma$ is not a choice: an additional algebraic field which removes the deformation upon elimination must couple linearly to $\p\cdot h'$, and it is therefore a traceless tensor of rank $s-3$. This is consistent with the fact, shown in Section~\ref{sec:pbf}, that the deformation only affects the sectors of spin $j\leq s-3$. Second, the argument establishes existence, not minimality. Minimality was proven for $s=3$, in the number of auxiliary components, while for $s=4$ we showed that the ADY content alone is not sufficient and that, within the branches we followed, one additional algebraic vector is needed.
\section{Massive completions of the Skvortsov--Vasiliev model}\label{sec:sv}

 We now turn to the second line of Table~\ref{tab:models}. In contrast with the Fronsdal model, whose massive completions are provided by the SH and ADY constructions, massive versions of the SV model have received much less attention. For $s=2$ the SV model reduces to the linearized WTDiff theory, and its massive versions were obtained in \cite{bfh} by dimensional reduction. For arbitrary $s$, a Stückelberg-like description of a massive field based on the Maxwell-like Lagrangians of \cite{cf}, that is, on the traceless formulation of the SV model, was given in \cite{bcps}, with traceless auxiliary fields of every rank $0,\dots,s-1$ and constrained gauge parameters. Here we follow a different route: we keep the SV Lagrangian of the double traceless field $h$ together with its Weyl symmetry, add only the auxiliary fields needed to remove the lower-spin modes, and show that, within a general ansatz, this auxiliary sector is fixed (for $s=4$, within the restriction of the ansatz specified in Section~\ref{sec:sv4}). The extension to arbitrary $s$, and to the PBSV models, is given in Section~\ref{sec:pbsv}. 

\subsection{Spin-3}\label{sec:sv3}
For $s=3$ the point (\ref{svpoint}) gives $a^*_3=2/(D+2)$, $b^*_3=(D+4)/(D+2)^2$ and $c^*_3=-2(D-2)/(D+2)^2$, and the SV Lagrangian reads
\be
\mathcal{L}_{SV}(h) = \frac12\, h\,\Box\, h + \frac32\,(\p\cdot h)^2 + \frac{6}{D+2}\, h'\cdot\p\cdot\p\cdot h - \frac{3(D+4)}{2(D+2)^2}\, h'\,\Box\, h' - \frac{3(D-2)}{2(D+2)^2}\,(\p\cdot h')^2\, . \label{lsv3}
\ee
Our construction rests on a simple observation. The traceless part of $h$,
\be
\hat h_{\mu\nu\rho} = h_{\mu\nu\rho} - \frac{1}{D+2}\left(\eta_{\mu\nu}\,h'_\rho+\eta_{\mu\rho}\,h'_\nu+\eta_{\nu\rho}\,h'_\mu\right) , \label{hhat}
\ee
is inert under the Weyl transformations of the SV model (Table~\ref{tab:models}). The Weyl invariance of (\ref{lsv3}) then implies that it can be written in terms of $\hat h$ alone, and since all the trace terms vanish on a traceless field, up to total derivatives,
\be
\mathcal{L}_{SV}(\hat h) = \frac12\,\hat h\,\Box\,\hat h + \frac32\,(\p\cdot\hat h)^2\, . \label{svfronsdal}
\ee
The SV model is thus the Fronsdal model in which the trace has been turned into a pure gauge mode. This identifies precisely what is missing with respect to the ADY construction: there, the trace $h'$ plays the role of the rank-$(s-2)$ auxiliary field of SH, while here it is gauge, and its role must be taken over by an independent field.

In the ADY model (\ref{dy3}) let us call $H$ its Fronsdal field. There, the situation is the opposite: its trace $H'$ is not gauge, but the vector auxiliary field of SH. Decomposing $H$ into its traceless part and its trace, the terms of $\mathcal{L}_F(H)$ which contain only the traceless part have, by (\ref{svfronsdal}), exactly the form of the SV Lagrangian. We therefore identify the traceless part of $H$ with $\hat h$, and rename its trace as an independent vector $A$:
\be
H_{\mu\nu\rho} = \hat h_{\mu\nu\rho} + \frac{1}{D+2}\left(\eta_{\mu\nu}A_\rho+\eta_{\mu\rho}A_\nu+\eta_{\nu\rho}A_\mu\right)\, , \qquad H' = A\, . \label{Hparam}
\ee
Substituting (\ref{Hparam}) into (\ref{dy3}), we obtain
\bea
\mathcal{L}_{SVm} &=& \mathcal{L}_{SV}(\hat h) - \frac{m^2}{2}\,\hat h\cdot\hat h + \frac{3D}{D+2}\, A\cdot\p\cdot\p\cdot\hat h - \frac{3D(D+1)}{2(D+2)^2}\, A\cdot\Box A + \frac{3D(D-2)}{4(D+2)^2}\,(\p\cdot A)^2 \nn\\
&+& \frac{3(D+1)}{2(D+2)}\, m^2\, A\cdot A + m\,\vf\,\p\cdot A - \frac{\beta}{2}\,\vf\,\Box\,\vf + \frac{\gamma}{2}\, m^2\vf^2\, , \label{lsvm}
\eea
with $\beta$ and $\gamma$ given in (\ref{betagamma}). Treated as an independent traceless field, $\hat h$ is inert, and (\ref{lsvm}) is simply the ADY Lagrangian in the variables $(\hat h,A,\vf)$, with no gauge symmetry. 

Starting from (\ref{lsvm}),  we can return to the original field $h$ by means of (\ref{hhat}). Since $\mathcal{L}_{SV}(\hat h)=\mathcal{L}_{SV}(h)$ up to total derivatives, this gives
\bea
\mathcal{L}_{SVm} &=& \mathcal{L}_{SV}(h) - \frac{m^2}{2}\left(h\cdot h - \frac{3}{D+2}\,h'\cdot h'\right) \nn\\
&+& \frac{3D}{D+2}\, A\cdot\p\cdot\p\cdot h - \frac{3D}{(D+2)^2}\, A\cdot\left(\Box h'+2\,\p\,\p\cdot h'\right) \nn\\
&-& \frac{3D(D+1)}{2(D+2)^2}\, A\cdot\Box A + \frac{3D(D-2)}{4(D+2)^2}\,(\p\cdot A)^2 + \frac{3(D+1)}{2(D+2)}\, m^2\, A\cdot A \nn\\
&+& m\,\vf\,\p\cdot A - \frac{\beta}{2}\,\vf\,\Box\,\vf + \frac{\gamma}{2}\, m^2\vf^2\, . \label{lsvmh}
\eea
Equivalently, (\ref{lsvmh}) follows from (\ref{dy3}) (in terms of $H$) by the single substitution
\be
H_{\mu\nu\rho}=h_{\mu\nu\rho}+\frac{1}{D+2}\Big(\eta_{\mu\nu}(A-h')_\rho+\eta_{\mu\rho}(A-h')_\nu+\eta_{\nu\rho}(A-h')_\mu\Big)\, ,
\ee
in which the trace of $h$ is replaced by $A$. This map is onto but not one-to-one: $(h,A)$ and $(h+\eta\,\sigma,A)$, with an arbitrary vector $\sigma$, give the same $H$. Hence (\ref{lsvmh}) is invariant, for any $m$, under the Weyl symmetry of the massless SV model,
\be
\delta h_{\mu\nu\rho} = \eta_{\mu\nu}\,\sigma_\rho+\eta_{\mu\rho}\,\sigma_\nu+\eta_{\nu\rho}\,\sigma_\mu\, , \qquad \delta A_\mu = \delta\vf=0\, , \label{weylm}
\ee
while the transverse diffeomorphisms are broken by the mass term. Both (\ref{lsvm}) and (\ref{lsvmh}) describe the same particle content, but only the latter is a theory of the SV field $h$, with the kinetic term and the gauge symmetry of the massless SV model, and it is the one we analyse below.

The corresponding Noether identity of (\ref{lsvmh}),
\be
E'^{\,\rho}_h\equiv0\, , \label{noetherw}
\ee
makes the matrices of the spin-1 and spin-0 sectors degenerate. As in the computation of a propagator, we add a gauge-fixing term,
\be
\mathcal{L}_{gf} = -\frac{\theta\, m^2}{2}\, h'\cdot h'\, , \qquad \theta\neq0\, , \label{lgf}
\ee
which simply replaces the coefficient $-\frac{3}{D+2}$ of $h'\cdot h'$ in the mass term by $-\frac{3}{D+2}+\theta$. Since (\ref{weylm}) shifts the trace algebraically, $\delta h'=(D+2)\,\sigma$, the gauge $h'=0$ is complete and could equally be imposed directly in the action \cite{motohashi}; with (\ref{lgf}).

The analysis of $\mathcal{L}_{SVm}+\mathcal{L}_{gf}$ then proceeds as for the PBF model. In the spin-3 sector all the trace and divergence terms drop out, together with the couplings to $A$ and $\vf$, and
\be
(\Box - m^2)\, h^{TT}_{\mu\nu\rho} = 0\, . \label{s3}
\ee
In the spin-2 sector, the Noether identity of the transverse diffeomorphisms of the massless SV model, $(\p\cdot E_h)^{TT}\equiv0$, receives no contribution from $A$ or from (\ref{lgf}), and the massive theory gives the algebraic constraint
\be
(\p\cdot E_h)^{TT}_{\nu\rho} = -\,m^2\,(\p\cdot h)^{TT}_{\nu\rho} = 0\, . \label{s2}
\ee
The spin-1 sector contains the transverse parts of $h'$, $\p\cdot\p\cdot h$ and $A$. Taking as rows the transverse parts of $E_h'$, $\p\cdot\p\cdot E_h$ and $E_A$, we find
\be
M^{(1)}=\begin{pmatrix}
	-\frac{\theta(D+2)}{3}\,m^2 & 0 & 0 \\[2pt]
	\frac{\Box\,[(D+2)m^2+D\Box]}{(D+2)^2}-\frac{\theta}{3}\,m^2\Box & -m^2-\frac{D}{D+2}\,\Box & \frac{D(D+1)}{(D+2)^2}\,\Box^2 \\[2pt]
	-\frac{3D}{(D+2)^2}\,\Box & \frac{3D}{D+2} & \frac{3(D+1)[(D+2)m^2-D\Box]}{(D+2)^2}
\end{pmatrix}, \label{s1matrix}
\ee
whose determinant is
\be
\det M^{(1)} = \theta\,(D+1)\, m^6\, . \label{s1det}
\ee
The constancy of the determinant results from a cancellation of the $\Box^2$ terms in the lower $2\times2$ block, which is the Weyl-invariant part of the sector. The spin-0 sector contains $\p\cdot\p\cdot\p\cdot h$, $\p\cdot h'$, $\p\cdot A$ and $\varphi$. Taking $\p\cdot E'_h$, $\p\cdot\p\cdot\p\cdot E_h$, $\p\cdot E_A$ and $E_\varphi$ as rows, and $\p\cdot\p\cdot\p\cdot h$, $\p\cdot h'$, $\p\cdot A$ and $\varphi$ as columns, we find
{\setlength{\arraycolsep}{3pt}
	\be
	M^{(0)}=\begin{pmatrix}
		0 & -\frac{\theta(D+2)}{3}\,m^2 & 0 & 0 \\[2pt]
		-m^2-\frac{2(D-1)}{D+2}\,\Box & \frac{3\Box\,[(D+2)m^2+2(D-1)\Box]}{(D+2)^2}-\theta\,m^2\Box & \frac{3D(D-1)}{(D+2)^2}\,\Box^2 & 0 \\[2pt]
		\frac{3D}{D+2} & -\frac{9D}{(D+2)^2}\,\Box & \frac{3(D+1)}{D+2}\,m^2-\frac{9D^2}{2(D+2)^2}\,\Box & -m\,\Box \\[2pt]
		0 & 0 & m & \gamma\,m^2-\beta\,\Box
	\end{pmatrix}, \label{s0matrix}
	\ee}
whose determinant is
\be
\det M^{(0)} = -\,\frac{2\,\theta\, D^2(D+1)}{(D-2)^2}\, m^8\, . \label{s0det}
\ee
At $\theta=0$ the first row vanishes, as the divergence of (\ref{noetherw}), and the first two columns become proportional, reflecting the gauge direction $(\delta\,\p\cdot\p\cdot\p\cdot h,\,\delta\,\p\cdot h')\propto(3\Box,\,D+2)$. The gauge-fixing term lifts this degeneracy, and $\det M^{(0)}$ is $\frac{\theta(D+2)}{3}m^2$ times the determinant of the Weyl-invariant $3\times3$ block, $-\frac{6D^2(D+1)}{(D+2)(D-2)^2}\,m^6$. Hence $\mathcal{L}_{SVm}$ propagates only the spin-3 modes, with mass $m$. The Weyl symmetry survives the mass, which is equivalent to the statement that $h'$ does not appear in $\mathcal{L}_{SVm}$ once it is written in terms of the Weyl-inert field $\hat h$, while the Noether identity of the transverse diffeomorphisms is replaced by the constraint (\ref{s2}), and the role played by the trace equation in the ADY model is taken over by the equation of $A$.

\label{sec:svunique}Since (\ref{lsvm}) was obtained by rewriting a known model, it is natural to ask whether other auxiliary sectors are possible. We keep the SV kinetic term for $h$ and consider the most general local Lagrangian with at most two derivatives, quadratic in $h$, $A$ and $\vf$:
\bea
\mathcal{L} &=& \mathcal{L}_{SV}(h) - \frac{m^2}{2}\left(h\cdot h + \alpha\, h'\cdot h'\right) + q_1\, A\cdot\p\cdot\p\cdot h + q_2\, A\cdot\Box h' + q_3\, A\cdot\p\,\p\cdot h' \nn\\
&+& q_4\, A\cdot\Box A + q_5\,(\p\cdot A)^2 + q_6\, m^2 A\cdot A + q_7\, m^2 A\cdot h' \nn\\
&+& q_8\, m\,\vf\,\p\cdot A + q_9\, m\,\vf\,\p\cdot h' + q_{10}\,\vf\,\Box\,\vf + q_{11}\, m^2\vf^2\, , \label{lsvansatz}
\eea
where the normalization of $A$ is fixed by $q_1=\frac{3D}{D+2}$; no other terms are allowed, since $A^\mu\p^\nu h_{\mu\nu\rho}$ leaves a free index and $h'\cdot\p\vf$ is equivalent to $\vf\,\p\cdot h'$. The spin-3 and spin-2 sectors are as in (\ref{s3}) and (\ref{s2}) for any choice of the coefficients, and the conditions come from the spin-1 and spin-0 sectors, which we solve in cascade as in Section~\ref{sec:spin3}. The scalar does not enter the spin-1 sector, whose determinant is a nonzero constant for $\alpha\neq-\frac{3}{D+2}$ if and only if
\be
q_2=-\frac{3D}{(D+2)^2}\, , \qquad q_4=-\frac{3D(D+1)}{2(D+2)^2}\, , \label{cascade1}
\ee
together with a relation fixing $q_6$. For $q_{10}\neq0$, the spin-0 sector then gives
\be
q_3=-\frac{6D}{(D+2)^2}\, , \qquad q_5=\frac{3D(D-2)}{4(D+2)^2}\, , \label{cascade0}
\ee
and two relations fixing $q_{10}$ and $q_{11}$. In the branch $q_7=q_9=0$ the solution is
\be
q_{10}=-\frac{\beta}{2}\,q_8^2\, , \qquad q_{11}=\frac{\gamma}{2}\,q_8^2\, , \qquad q_6=\frac{3(D+1)}{2(D+2)}\, , \qquad \alpha=-\frac{3}{D+2}+\theta\, , \quad \theta\neq0\, , \label{cascadesol}
\ee
and the determinants of the spin-1 and spin-0 sectors are (\ref{s1det}) and (\ref{s0det}). These are exactly the coefficients of (\ref{lsvm}), with $q_8$ the normalization of $\vf$, and since $h\cdot h-\frac{3}{D+2}h'\cdot h'=\hat h\cdot\hat h$, the mass term splits as $-\frac{m^2}{2}\hat h\cdot\hat h-\frac{\theta m^2}{2}h'\cdot h'$. At $\theta=0$ the determinants vanish identically because of the Weyl symmetry. The general solution, with $q_7$ and $q_9$ free, follows from this one by the Weyl transformation (\ref{weylm}) with $\sigma=\lambda A+\mu\,\p\vf/m$, which leaves $\hat h$ invariant and only modifies (\ref{lgf}); the determinants remain nonzero provided $q_7q_9+\theta\,q_8\neq0$. Within the ansatz (\ref{lsvansatz}) with $q_{10}\neq0$, the massive spin-3 SV model is therefore unique up to the normalization of $\vf$ and the choice of Weyl gauge.

\subsection{Spin-4}\label{sec:sv4}

For $s=4$ the point (\ref{svpoint}) gives $a^*_4=2/(D+4)$, $b^*_4=(D+6)/(D+4)^2$ and $c^*_4=-2D/(D+4)^2$, and
\be
\mathcal{L}_{SV}^{(4)}(h) = \frac12\, h\,\Box\, h + 2\,(\p\cdot h)^2 + \frac{12}{D+4}\, h'\cdot\p\cdot\p\cdot h - \frac{3(D+6)}{(D+4)^2}\, h'\cdot\Box\, h' - \frac{6D}{(D+4)^2}\,(\p\cdot h')^2\, , \label{lsv4}
\ee
with $h''=0$. It is invariant under $\delta h=\p\Lambda$, with $\Lambda$ traceless and transverse, and under Weyl transformations $\delta h=\eta\,\bar\sigma$, with $\bar\sigma'=0$ and $\delta h'=(D+4)\,\bar\sigma$. As for $s=3$, the traceless part
\be
\hat h = h - \frac{1}{D+4}\,\eta\, h' \label{hhat4}
\ee
is Weyl inert, and
\be
 \mathcal{L}_F^{(4)}(\hat h) = \frac12\,\hat h\,\Box\,\hat h + 2\,(\p\cdot\hat h)^2\, . \label{svfronsdal4}
\ee
Starting from the ADY Lagrangian (\ref{dy4}), we parametrize the Fronsdal field as
\be
H = \hat h + \frac{1}{D+4}\,\eta\, A\, , \qquad A' = 0\, , \label{Hparam4}
\ee
where $A$ is now a symmetric traceless rank-2 tensor, so that $H'=A$, and we rename $a_1\to B$ and $a_0\to\vf$. Using (\ref{svfronsdal4}) and $H\cdot H-6\,H'\cdot H'=\hat h\cdot\hat h-\frac{6(D+3)}{D+4}\,A\cdot A$, we obtain
\bea
\mathcal{L}_{SVm}^{(4)} &=& \mathcal{L}_{SV}^{(4)}(\hat h) - \frac{m^2}{2}\,\hat h\cdot\hat h + \frac{6(D+2)}{D+4}\, A\cdot\p\cdot\p\cdot\hat h - \frac{3(D+2)(D+3)}{(D+4)^2}\, A\cdot\Box A \nn\\
&+& \frac{3D(D+2)}{(D+4)^2}\,(\p\cdot A)^2 + \frac{3(D+3)}{D+4}\, m^2\, A\cdot A + 12\, m\, B\cdot\p\cdot A \nn\\
&+& \frac32\Big[ -\beta_4\, B\cdot\Box\, B + \gamma_4\, (\p\cdot B)^2 - \sigma_4\, \vf\,\Box\, \vf + m^2\left(\tau_4\, B\cdot B + \rho_4\, \vf^2\right) - 2\, m\, (\p\cdot B)\, \vf \Big] ,\nn\\
 \label{lsvm4}
\eea
with the coefficients (\ref{dy4coef}). Being related to the ADY model by a local and invertible change of variables, (\ref{lsvm4}) describes a single massive spin-4 particle, with an auxiliary sector made of a traceless rank-2 tensor, a vector and a scalar. Writing $\hat h$ back in terms of $h$ through (\ref{hhat4}), or equivalently substituting $H=h+\frac{1}{D+4}\,\eta\,(A-h')$ directly in (\ref{dy4}), we obtain
\bea
\mathcal{L}_{SVm}^{(4)} &=& \mathcal{L}_{SV}^{(4)}(h) - \frac{m^2}{2}\Big(h\cdot h-\frac{6}{D+4}\,h'\cdot h'\Big) + \frac{6(D+2)}{D+4}\, A\cdot\p\cdot\p\cdot h \nn\\
&-& \frac{6(D+2)}{(D+4)^2}\Big(A\cdot\Box\, h' + 4\, A^{\mu\nu}\p_\mu\p^\rho h'_{\nu\rho}\Big) - \frac{3(D+2)(D+3)}{(D+4)^2}\, A\cdot\Box A \nn\\
&+& \frac{3D(D+2)}{(D+4)^2}\,(\p\cdot A)^2 + \frac{3(D+3)}{D+4}\, m^2\, A\cdot A + 12\, m\, B\cdot\p\cdot A \nn\\
&+& \frac32\Big[ -\beta_4\, B\cdot\Box\, B + \gamma_4\, (\p\cdot B)^2 - \sigma_4\, \vf\,\Box\, \vf + m^2\left(\tau_4\, B\cdot B + \rho_4\, \vf^2\right) - 2\, m\, (\p\cdot B)\, \vf \Big] .\nn\\ \label{lsvm4h}
\eea
Since $h$ enters (\ref{lsvm4h}) only through $\hat h$, the Lagrangian is invariant, for any $m$, under
\be
\delta h = \eta\,\bar\sigma\, , \qquad \delta A = \delta B = \delta\vf = 0\, , \qquad \bar\sigma'=0\, , \label{weylm4}
\ee
and, as for $s=3$, we fix this symmetry with
\be
\mathcal{L}_{gf} = -\frac{\theta\, m^2}{2}\, h'\cdot h'\, , \qquad \theta\neq0\, , \label{lgf4}
\ee
which replaces the coefficient $-\frac{6}{D+4}$ of $h'\cdot h'$ in the mass term by $-\frac{6}{D+4}+\theta$.

 The equation of motion of $h$ is projected onto its double-traceless part, $E_h^{DT}=E_h-\frac{1}{D(D+2)}\,\eta\,\eta\, E_h''$, and that of $A$ onto its traceless part. In the spin-4 sector only $\mathcal{L}_{SV}$ and the mass term contribute, and
 \be
 (\Box-m^2)\, h^{TT} = 0\, , \label{sv4s4}
 \ee
 while in the spin-3 sector the transverse diffeomorphism identity of $\mathcal{L}_{SV}$ reduces the equations to the algebraic constraint
 \be
 (\p\cdot E_h)^{TT} = -\,m^2\,(\p\cdot h)^{TT} = 0\, . \label{sv4s3}
 \ee
 The spin-2 sector contains $h'^{TT}$, $(\p\cdot\p\cdot h)^{TT}$ and $A^{TT}$, and its rows are the trace and the double divergence of $E_h^{DT}$ and $E_A$. One more divergence of these equations, together with $E_B$, gives the spin-1 sector, with $(\p\cdot h')^T$, $(\p\cdot\p\cdot\p\cdot h)^T$, $(\p\cdot A)^T$ and $B^T$, and a further one, together with $E_\vf$, the spin-0 sector, with $\p\cdot\p\cdot h'$, $\p\cdot\p\cdot\p\cdot\p\cdot h$, $\p\cdot\p\cdot A$, $\p\cdot B$ and $\vf$. In each sector the row coming from the trace of $E_h$ has the single entry $-\frac{(D+4)\theta}{6}\,m^2$, as the projection of the Weyl identity, and the determinants are nonzero constants which vanish at $\theta=0$,
 \bea
 &&\det M^{(2)}=(D+3)\,\theta\, m^6\, , \qquad \det M^{(1)}=-\frac{72\,(D+2)^2(D+3)}{D^2}\,\theta\, m^8\, , \nn\\
 &&\det M^{(0)} = -\frac{54\, D(D+1)(D+2)}{(D-2)^2}\,\theta\, m^{10}\, . \label{sv4dets}
 \eea
 Hence $\mathcal{L}_{SVm}^{(4)}+\mathcal{L}_{gf}$ propagates only the spin-4 modes, with mass $m$.

As for $s=3$, the auxiliary sector of (\ref{lsvm4h}) is fixed by the absence of lower-spin modes. We keep the SV kinetic term for $h$ and consider the most general local Lagrangian with at most two derivatives, quadratic in $h$, $A$, $B$ and $\vf$:
\bea
\mathcal{L} &=& \mathcal{L}_{SV}^{(4)}(h) - \frac{m^2}{2}\left(h\cdot h + \alpha\, h'\cdot h'\right) + q_1\, A\cdot\p\cdot\p\cdot h + q_2\, A\cdot\Box\, h' + q_3\, A^{\mu\nu}\p_\mu\p^\rho h'_{\nu\rho} \nn\\
&+& q_4\, A\cdot\Box A + q_5\,(\p\cdot A)^2 + q_6\, m^2 A\cdot A + q_7\, m^2 A\cdot h' \nn\\
&+& q_8\, m\,\vf\,\p\cdot B + q_9\,\vf\,\p\cdot\p\cdot h' + q_{10}\,\vf\,\Box\,\vf + q_{11}\, m^2\vf^2 + q_{17}\,\vf\,\p\cdot\p\cdot A \nn\\
&+& q_{12}\, m\, B\cdot\p\cdot h' + q_{13}\, m\, B\cdot\p\cdot A + q_{14}\, B\cdot\Box B + q_{15}\,(\p\cdot B)^2 + q_{16}\, m^2 B\cdot B\, . \label{lsvansatz4}
\eea
The coefficients $q_1$, $q_{13}$ and $q_8$ fix the normalizations of $A$, $B$ and $\vf$, and we take $q_1=\frac{6(D+2)}{D+4}$ and $q_{13}=12$, as in (\ref{lsvm4h}). We set $q_7=q_9=q_{12}=0$, so that $h'$ couples to the auxiliary fields only through the derivative couplings $q_2$ and $q_3$ to $A$. Nonzero $q_7$ and $q_{12}$ are generated from the solution below by the Weyl transformation (\ref{weylm4}) with $\bar\sigma=\lambda A+\frac{\mu}{m}\big(\p B-\frac{2}{D}\,\eta\,\p\cdot B\big)$, which leaves $\hat h$ invariant and, acting on (\ref{lgf4}), gives $q_7=-(D+4)\,\theta\lambda$ and $q_{12}=2(D+4)\,\theta\mu$, together with terms quadratic in $A$ and $B$ which shift $q_6$, $q_{13}$, $q_{14}$ and $q_{15}$. The coupling $q_9$ is not generated in this way, and $q_9=0$ is a restriction of the ansatz. The sectors of spin 4 and 3 are as in (\ref{sv4s4}) and (\ref{sv4s3}) for any choice of the remaining coefficients, and, with $G\equiv6+(D+4)\alpha$, the matrices of the lower-spin sectors, with rows and columns ordered as above (\ref{sv4dets}), are
{\setlength{\arraycolsep}{4pt}\allowdisplaybreaks
	\begin{align}
		M^{(2)}&=\begin{pmatrix}
			-\frac{G}{6}\,m^2 & 0 & \frac{P_2}{6}\,\Box \\[4pt]
			b_2 & e_2 & \frac{q_1+q_2}{6}\,\Box^2 \\[4pt]
			q_2\,\Box & q_1 & 2\,(q_6\,m^2+q_4\,\Box)
		\end{pmatrix}, \label{M2ans4}\\[8pt]
		M^{(1)}&=\begin{pmatrix}
			0 & -\frac{G}{6}\,m^2 & \frac{P_1}{12}\,\Box & 0 \\[4pt]
			e_1 & b_1 & \frac{2q_1+2q_2+q_3}{4}\,\Box^2 & 0 \\[4pt]
			q_1 & \big(q_2+\frac{q_3}{2}\big)\Box & 2q_6\,m^2+(2q_4-q_5)\,\Box & -\frac{q_{13}}{2}\,m\,\Box \\[4pt]
			0 & 0 & q_{13}\,m & 2\,(q_{16}\,m^2+q_{14}\,\Box)
		\end{pmatrix}, \label{M1ans4}\\[8pt]
		M^{(0)}&=\begin{pmatrix}
			0 & -\frac{G}{6}\,m^2 & \frac{P_0}{6D}\,\Box & 0 & 0 \\[4pt]
			e_0 & b_0 & \frac{R}{D(D+2)}\,\Box^2 & 0 & 0 \\[4pt]
			q_1 & s\,\Box & 2\,(q_6\,m^2+t\,\Box) & -\omega\,q_{13}\,m\,\Box & \omega\,q_{17}\,\Box^2 \\[4pt]
			0 & 0 & q_{13}\,m & 2\,(q_{16}\,m^2+\Delta\,\Box) & -q_8\,m\,\Box \\[4pt]
			0 & 0 & q_{17} & q_8\,m & 2\,(q_{11}\,m^2+q_{10}\,\Box)
		\end{pmatrix}, \label{M0ans4}
\end{align}}
The entries $e_j$ and $b_j$ come only from $\mathcal{L}_{SV}^{(4)}$ and from the mass term, and do not depend on the auxiliary sector,
\begin{align*}
	e_j&=-m^2-c_j\,\Box\, , & c_j&=\frac{1}{D+4}\,\big(D+2,\ 2(D+1),\ 3D\big)\, ,\\
	b_j&=-n_j\,\alpha\,m^2\Box+d_j\,\Box^2\, , & n_j&=\Big(\frac16,\ \frac12,\ 1\Big)\, , \qquad d_j=\frac{1}{(D+4)^2}\,\big(D+2,\ 6(D+1),\ 18D\big)\, ,
\end{align*}
for $j=2,1,0$. The couplings between $h$ and $A$ enter through the combinations
\begin{align*}
	P_2&=q_1+(D+4)\,q_2\, , \qquad P_1=6q_1+(D+4)(2q_2+q_3)\, ,\\
	P_0&=(5D-4)\,q_1+D(D+4)\,q_2+(D-1)(D+4)\,q_3\, ,\\
	R&=(D^2+2D-2)\,q_1+D(D+2)\,q_2+(D-1)(D+2)\,q_3\, , \qquad s=\frac{D\,q_2+(D-1)\,q_3-q_1}{D}\, ,
\end{align*}
and the kinetic coefficients of $A$ and $B$ through $t=q_4-\omega\,q_5$ and $\Delta=q_{14}-q_{15}$, with $\omega=\frac{D-1}{D}$. The determinants are polynomials of degree 3, 4 and 4 in $\Box$. We require them to be nonzero constants and solve in cascade, starting from the highest power of $\Box$ and from the spin-2 sector, which involves the fewest coefficients. At each step the condition factorizes, and the factors which cannot vanish are $G$, the normalizations and the kinetic coefficients $q_{10}$, $q_{14}$ and $\Delta$. In the spin-0 sector the coefficient of $\Box^4$ is proportional to $\Delta\,q_{17}^2$, so that $q_{17}=0$ is required. The unique solution is
\begin{alignat}{3}
	q_2 &= -\frac{6(D+2)}{(D+4)^2}\, , &\qquad q_3 &= -\frac{24(D+2)}{(D+4)^2}\, , &\qquad q_4 &= -\frac{3(D+2)(D+3)}{(D+4)^2}\, , \nonumber\\
	q_5 &= \frac{3D(D+2)}{(D+4)^2}\, , &\qquad q_6 &= \frac{3(D+3)}{D+4}\, , &\qquad q_{17} &= 0\, , \nonumber\\
	q_{14} &= -\frac32\,\beta_4\, , &\qquad q_{15} &= \frac32\,\gamma_4\, , &\qquad q_{16} &= \frac32\,\tau_4\, , \nonumber\\
	q_{10} &= -\frac{\sigma_4}{6}\,q_8^2\, , &\qquad q_{11} &= \frac{\rho_4}{6}\,q_8^2\, . && \label{sol4sv}
\end{alignat}
\no with $\alpha=-\frac{6}{D+4}+\theta$, $\theta\neq0$, and the determinants are then (\ref{sv4dets}) multiplied by $q_8^2/9$ in the spin-0 sector. For $q_8=-3$ these are exactly the coefficients of (\ref{lsvm4h}) together with the gauge-fixing term (\ref{lgf4}). In particular $P_2=P_1=P_0=0$, so that the first row of each matrix reduces to the gauge-fixing entry, as for $s=3$. Within the ansatz (\ref{lsvansatz4}) restricted by $q_7=q_9=q_{12}=0$, the massive spin-4 SV model is therefore unique up to the normalizations of the auxiliary fields and the choice of the gauge-fixing parameter $\theta$, and at $\theta=0$ the determinants vanish identically because of the Weyl symmetry.

\section{Massive PBSV models: arbitrary spin}\label{sec:pbsv}

The PBSV model (\ref{lpbsv}) stands to the SV model as the PBF model stands to Fronsdal's, and its structure is much simpler. Let
\be
\hat h = h-\frac{1}{D+2(s-2)}\,\eta\, h'\, , \label{hhats}
\ee
be the traceless part of $h$, which reduces to (\ref{hhat}) and (\ref{hhat4}) for $s=3,4$. As in those cases, $\hat h$ is inert under the Weyl transformations, so that the SV Lagrangian depends on $h$ only through $\hat h$, and $\mathcal{L}^{(s)}_{SV}(h)=\mathcal{L}^{(s)}_{SV}(\hat h)=\mathcal{L}^{(s)}_F(\hat h)$, the last equality because every $h'$-dependent term of (\ref{labc}) vanishes on a traceless field. The deformation, instead, depends only on $h'$, and since $h\to(\hat h,h')$ is a local and invertible change of variables, (\ref{lpbsv}) splits into two decoupled sectors,
\be
\mathcal{L}^{(s)}_{PBSV}(c)=\mathcal{L}^{(s)}_F(\hat h)+\frac{c-c^*_s}{4\kappa}\,(\p\cdot h')^2\, , \label{pbsvsplit}
\ee
with $\kappa$ as in (\ref{lsgen}). The first sector is the traceless formulation of the SV model \cite{sv}. The second one is invariant under $\delta h'=[D+2(s-2)]\,\bar\sigma^T$, and its equation of motion is the traceless part of $\p\,(\p\cdot h')=0$, which has the form $\p\psi+\eta\chi=0$. For fields which vanish at infinity, such equations imply $\psi=\chi=0$ \cite{dmmb}, and
\be
\p\cdot h'=0\, , \label{dhpzero}
\ee
while the transverse part of $h'$ is pure gauge. The sector of $h'$ carries no propagating modes, and its only effect is the contact term in the coupling to sources found in \cite{dmmb}.

The construction of Section~\ref{sec:sv} is not restricted to $s=3,4$. For any $s$, the parametrization $H=\hat h+\frac{1}{D+2(s-2)}\,\eta\,A$, with $H'=A$, and $A$ a traceless tensor of rank $s-2$, is a local and invertible change of variables, and substituted into the ADY Lagrangian (\ref{ladys}) it gives a massive SV model, which is explicitly written as:
\bea
\mathcal{L}^{(s)}_{SVm} &=& \mathcal{L}^{(s)}_{SV}(\hat h) - \frac{m^2}{2}\,\hat h\cdot\hat h + \frac{s(s-1)(D+2s-6)}{2(D+2s-4)}\, A\cdot\p\cdot\p\cdot\hat h \nn\\
&-& \frac{s(s-1)(D+2s-5)(D+2s-6)}{4(D+2s-4)^2}\, A\cdot\Box A + \frac{s(s-1)(D+2s-5)}{4(D+2s-4)}\, m^2\, A\cdot A \nn\\
&+& \frac{s(s-1)(s-2)(D+2s-6)(D+2s-8)}{8(D+2s-4)^2}\,(\p\cdot A)^2 \nn\\
&+& g_s\, m\, a_{s-3}\cdot\p\cdot A + \mathcal{L}_{aux}(a_{s-3},a_{s-4})\, , \label{lsvmsexp}
\eea
which reduces to (\ref{lsvm}) and (\ref{lsvm4}) for $s=3,4$.\footnote{We have also checked the coefficients of (\ref{lsvmsexp}) explicitly for $s=5$ and $s=6$ with xAct.} Its kinetic term for $h$ is $\mathcal{L}^{(s)}_{SV}(h)=\mathcal{L}^{(s)}_F(\hat h)$ and, being related to (\ref{ladys}) by a local and invertible change of variables up to the pure gauge trace mode, it describes a single massive spin-$s$ particle. Substituting (\ref{hhats}), it can be written in terms of the original field $h$, as we have done before for $s=3, 4$, but this form will not be needed here. The auxiliary sectors of (\ref{lsvm}) and (\ref{lsvm4}) were shown to be fixed within the ansatz of Section~\ref{sec:svunique} for $s=3$ and within that of Section~\ref{sec:sv4}, with $q_7=q_9=q_{12}=0$, for $s=4$. Because of (\ref{pbsvsplit}), the massive PBSV model is then obtained simply by replacing the SV kinetic term by the PBSV one,
\be
\mathcal{L}^{(s)}_{PBSVm}=\mathcal{L}^{(s)}_{SVm}\Big|_{\mathcal{L}_{SV}\to\mathcal{L}_{PBSV}(c)}=\mathcal{L}^{(s)}_{SVm}+\frac{c-c^*_s}{4\kappa}\,(\p\cdot h')^2\, , \label{lpbsvm}
\ee
and no auxiliary field beyond those of the massive SV model is needed. In the variables $(\hat h,h')$, (\ref{lpbsvm}) is the sum of the massive SV model, which depends on $\hat h$, $A$ and the auxiliary fields, and of the same sector of $h'$ as in the massless theory, which does not depend on $m$. For any value of $m$ it is invariant under
\be
\delta h=\eta\,\bar\sigma^T\, ,\qquad \p\cdot\bar\sigma^T=0\, , \label{pbsvmsym}
\ee
with $A$ and the auxiliary fields inert: as for the SV model, the mass breaks the transverse diffeomorphisms but preserves the Weyl symmetry of the massless theory, here in its transverse form.

The particle content follows at once. The spin-projection matrices of (\ref{lpbsvm}) are block diagonal in the variables $(\hat h,h')$. The block of $\hat h$, $A$ and the auxiliary fields is that of the massive SV model written in terms of the Weyl-inert field $\hat h$. It is related to that of ADY by the change of variables $H'=A$, $\hat h=H-\frac{1}{D+2(s-2)}\,\eta\,H'$, which is local, algebraic and triangular, so that in each spin sector both the variables and the equations are related by matrices with constant nonvanishing determinant. Hence, as for ADY, it has $\det M^{(s)}\propto\Box-m^2$ and constant determinants for $j<s$, with no need of gauge fixing. In the block of $h'$, the transverse and traceless part, of spin $s-2$, is pure gauge: the corresponding row of the spin-$(s-2)$ matrix vanishes identically, as the projection of the Noether identity of (\ref{pbsvmsym}), and the corresponding column is a null direction. The criterion (\ref{criterion}) is applied after removing this row and this column, that is, to the gauge invariant variables.\footnote{The algebraic gauge-fixing term (\ref{lgf}) used for the SV model is not appropriate here. It sets $h'=0$, while only the transverse part of $h'$ is pure gauge in (\ref{lpbsvm}). Its longitudinal part is not gauge, and the term $-\frac{\theta m^2}{2}h'\cdot h'$ changes its dynamics. For $s=3$, writing the longitudinal part of $h'$ as $\p\phi$, the equation of motion of $\phi$ becomes $\Box\big[\frac{c-c^*_3}{2\kappa}\,\Box+\theta m^2\big]\phi=0$, with a spurious pole at $\Box=-2\kappa\theta m^2/(c-c^*_3)$.} The components of $h'$ of spin $j\le s-3$ are set to zero by (\ref{dhpzero}). If they are described by $\p\cdot h'$, with the divergence of $E_{h'}$ as the corresponding row, the entries of the block of $h'$ carry the factor $c-c^*_s$ and, because of the divergence, a factor of $\Box$. For $s=3$, for instance, the spin-0 matrix of (\ref{lpbsvm}), with the rows and columns of (\ref{s0matrix}), has
\be
\det M^{(0)}=-\frac{3D^2(D+1)}{(D-2)^2}\,(c-c^*_3)\,\Box\, m^6\, ,
\ee
that is, $\frac{D+2}{2}(c-c^*_3)\,\Box$ times the Weyl-invariant block of the massive SV model. This factor of $\Box$ does not signal a propagating mode. The equation $E_{h'}=0$ gives (\ref{dhpzero}) directly, and every solution of (\ref{dhpzero}) is traceless, since $h''=0$, and transverse, hence of the form $h'=[D+2(s-2)]\,\bar\sigma^T$ and pure gauge under (\ref{pbsvmsym}). The zero at $\Box=0$ reflects this: in momentum space, at $k^2=0$ the longitudinal part of $h'$, which for $s=3$ is $\p\phi$, becomes transverse, and therefore pure gauge. Hence (\ref{lpbsvm}) describes a single massive spin-$s$ particle for every $c\neq c^*_s$.

The symmetry (\ref{pbsvmsym}) also fixes the sources. As in the massless model, it only requires the trace of the source to be orthogonal to every traceless and transverse tensor, that is, $T'=\eta\,\psi+\p\,\xi$, with $\xi$ a traceless tensor of rank $s-3$ ($T'_\mu=\p_\mu\xi$ for $s=3$), while no condition on its divergence is needed, since the transverse diffeomorphisms are broken by the mass. With $N=D+2(s-2)$, (\ref{hhats}) gives $h\cdot T=\hat h\cdot\hat T+\binom{s}{2}\frac{1}{N}\,h'\cdot T'$, where $\hat T$ is the traceless part of $T$. Since $h'$ is traceless, $h'\cdot\eta\,\psi=0$, and, up to a total derivative, $h'\cdot\p\,\xi=-(s-2)\,\xi\cdot\p\cdot h'$, so that the sector of $h'$ couples to the sources only through $-\frac{3}{N}\binom{s}{3}\,\xi\cdot\p\cdot h'$. Hence
\be
\bar T\,G^{-1}\,T=\big(\bar T\,G^{-1}\,T\big)_{ADY}\Big|_{\hat T}+\frac{\mathcal{C}_s}{c-c^*_s}\,\xi^2\, . \label{amppbsvm}
\ee
 The first term is the amplitude of the massive SV model, which coincides with the ADY one for traceless sources. The second one comes from the sector of $h'$, which is the same as in the massless model. Since $\frac{1}{4\kappa}=\frac34\binom{s}{3}$, this sector is $\frac34\binom{s}{3}\big[(c-c^*_s)\,(\p\cdot h')^2-\frac{4}{N}\,\xi\cdot\p\cdot h'\big]$, and the elimination of $\p\cdot h'$ gives
\be
\mathcal{C}_s=\frac{12}{[D+2(s-2)]^2}\binom{s}{3}\, , \label{cs}
\ee
with $\xi^2\equiv\xi\cdot\xi$. This reproduces the contact terms of the massless PBSV amplitudes, $\mathcal{C}_3=12/(D+2)^2$, eq.~(73) of \cite{dmmb}, and $\mathcal{C}_4=48/(D+4)^2$, eq.~(82), where $T'_{\mu\nu}=\eta_{\mu\nu}\psi+\p_\mu\xi_\nu+\p_\nu\xi_\mu$. Since the second term does not depend on $m$, $c$ survives the introduction of the mass, but only through a sector without particle content.

In the completion (\ref{lsgen}) of the PBF model, instead, $c$ does not survive, and the difference lies in the role of the trace. In the Fronsdal and ADY models $h'$ is the rank-$(s-2)$ auxiliary field of SH, and in (\ref{lsgen}) the deformation is neutralized by an algebraic field which undoes it; in the SV model the trace is pure gauge, its role is taken over by $A$, and the deformation acts on a decoupled sector. A completion of the PBF type is nevertheless also possible: adding to (\ref{lpbsvm}) the algebraic field $\sigma$ of (\ref{lsgen}), with $c-1\to c-c^*_s$, gives
\be
\mathcal{L}^{(s)}_{PBSVm}+(c-c^*_s)\Big[m\,\sigma\cdot\p\cdot h'+\kappa\, m^2\,\sigma\cdot\sigma\Big]=\mathcal{L}^{(s)}_{SVm}+(c-c^*_s)\,\kappa\, m^2\Big(\sigma+\frac{1}{2\kappa m}\,\p\cdot h'\Big)^2 , \label{pbsvsigma}
\ee
which is invariant under the full Weyl symmetry, with $\delta\sigma=-\frac{D+2(s-2)}{2\kappa m}\,\p\cdot\bar\sigma$, and from which the elimination of $\sigma$ gives back the massive SV model, with no dependence on $c$. Both (\ref{lpbsvm}) and (\ref{pbsvsigma}) describe a single massive spin-$s$ particle. The first one has one auxiliary field fewer and keeps the symmetry, the sources and the contact term of the massless PBSV model, while the second one removes $c$ altogether, as in the PBF case.
\section{Conclusion}\label{sec:conclusion}

We have studied the massive completions of two families of partially broken higher-spin models, both of which belong to the family of second order Lagrangians $\mathcal{L}^{(s)}(a,b,c)$ of \cite{pbf}: the PBF models, which deform the Fronsdal model, and the PBSV models, which deform the SV model. In both cases the deformation is the coefficient $c$ of the term $(\p\cdot h')^2$, and in both cases it affects the massless theory only through a contact term in the two-point amplitudes \cite{dmmb}. The question we have addressed is whether $c$ survives the introduction of a mass.

For the minimal massive completion of the PBF models the answer is negative. Working from first principles, with the most general local auxiliary sector with at most two derivatives and the spin projection criterion (\ref{criterion}), we have shown that for $s=3$ a single auxiliary scalar is never sufficient, while two scalars lead to a solution which is unique up to field redefinitions. One of the scalars is algebraic and its elimination exactly undoes the deformation, while the other one is the auxiliary scalar of the ADY model. For $s=4$ the same structure appears, with the ADY vector and scalar supplemented by an algebraic vector, and the construction extends to every $s$, Eq.~(\ref{lsgen}), with one algebraic traceless field $\sigma$ of rank $s-3$. Since $\sigma$ has a constant quadratic term and only couples to $h$, the Gaussian integration over it gives the ADY generating functional times a field independent factor. The correlation functions of $h$ and of the ADY auxiliary fields therefore coincide with those of the ADY model for every $m\neq0$, and hence with those of the SH model, whose unitarity in arbitrary dimension was established in \cite{SchimidtElias}; in particular they do not depend on $c$. Equivalently, in the regular form (\ref{square}) the whole dependence on $c$ is carried by the algebraic field $\sigma+\frac{1}{2\kappa m}\p\cdot h'$, which decouples from the other fields and from the sources, also in the Stückelberg formulation \cite{ady,zinoviev}.

The limit $m\to0$ of the minimal completion is therefore not continuous. At $m=0$ the Lagrangian (\ref{lsgen}) reduces to $\mathcal{L}_{PBF}^{(s)}(c)$ plus auxiliary fields which decouple from $h$, but for every $m\neq0$ its amplitudes are those of the ADY model and do not depend on $c$. For $s=3$, the sources allowed by the PBF symmetry, (\ref{sources}), are not conserved in the sense of the Fronsdal theory when $J\neq0$, since the traceless part of $\p\cdot T$ is then the traceless part of $\p\,\p J$, and their massive amplitude may contain terms singular in $m$. In any case its limit $m\to0$, when it exists, does not depend on $c$, and cannot reproduce the contact term $3J^2/(c-1)$ of (\ref{amp3}) for every $c$.

This is a property of the minimal completion rather than of the PBF model itself. In the Kaluza--Klein reduction of the spin-3 and spin-4 PBF models \cite{pbfkk}, the deformation is carried by the rank-$(s-2)$ Stückelberg field. Written in terms of gauge invariants, the reduced spin-3 model is the SH model plus $(c-1)$ times the square of an independent scalar invariant, which is the structure of (\ref{square}). In terms of the fundamental fields, however, the square contains the divergence of the Stückelberg vector, which survives in the massless limit as a spin-1 field mixed non-diagonally with the spin-3 one, and with sources coupled to it the massless limit reproduces the PBF amplitude, contact term included \cite{pbfkk}. The two completions have the same particle content and differ only in the sector which carries $c$: an algebraic field in the minimal completion, a Stückelberg gauge field in the Kaluza--Klein one. For $s=4$ the reduction produces two squares, of a vector and of a scalar, while the minimal completion requires only the vector.

For the SV line the situation is different. The SV Lagrangian is the Fronsdal Lagrangian of the traceless part of $h$, and its trace is pure gauge. Reinstating the trace as an independent traceless field of rank $s-2$, we have obtained a massive SV model for every $s$, related to the ADY model by a local and invertible change of variables, in which the Weyl-like symmetry survives the mass while the transverse diffeomorphisms are broken. Its correlation functions for traceless sources are those of the ADY model. For $s=3$, and for $s=4$ within an ansatz restricted by $q_7=q_9=q_{12}=0$, we have shown that its auxiliary sector is fixed by the absence of lower-spin modes, up to normalizations, mixing terms and the choice of Weyl gauge. A massive PBSV model is obtained with no auxiliary field beyond those of the massive SV model, since the deformation acts on the trace of $h$, which is decoupled from the massive SV sector. It keeps the transverse Weyl symmetry of the massless PBSV model, and its amplitude (\ref{amppbsvm}) is, for every $m\neq0$, the ADY one plus the same contact term as in the massless theory: in this case $c$ survives the mass, but only through a sector without particle content. Adding the algebraic field of the PBF construction, Eq.~(\ref{pbsvsigma}), restores the full Weyl symmetry and removes $c$, so that, as in the PBF case, whether $c$ survives is a matter of the chosen completion. The origin of the difference between (\ref{lsgen}) and (\ref{lpbsvm}) is the role of the trace, which is an essential auxiliary field in the Fronsdal line and a pure gauge mode in the SV line.

Several questions remain open. The first one is the minimality of the auxiliary content for general $s$, which we have conjectured for the PBF models; the analogous question for the SH and ADY models themselves was already left open in \cite{sh}. Second, for $s=4$ we have worked with $q_7=q_9=q_{12}=0$. Nonzero $q_7$ and $q_{12}$ are generated by Weyl transformations, as $q_7$ and $q_9$ are for $s=3$, but $q_9$ is not, and whether solutions with $q_9\neq0$ exist has not been determined. Third, the Kaluza--Klein reduction of the PBF models has been carried out for $s=3$ and $s=4$ in \cite{pbfkk}; its extension to arbitrary $s$, and to the PBSV models, would clarify how the sector which carries $c$ depends on the choice of massive completion.
	
	\section*{Acknowledgements}

We thank D.~Dalmazi and A.~L.~R.~dos~Santos for useful discussions. This work was supported by FAPESP under grant 2025/18083-7. We acknowledge the use of Claude (Anthropic), an AI assistant, to improve the English of the manuscript, to check statements, and to help with the writing of xAct (Wolfram Mathematica) and \LaTeX\ code. The authors take full responsibility for the content of this work.


\end{document}